\documentclass[iop,twocolumn,twocolappendix,numberedappendix,appendixfloats]{openjournal}

\usepackage{setspace}
\usepackage{apjfonts}

\usepackage{amsmath}
\usepackage[T1]{fontenc}

\usepackage{CJK}
\usepackage[hidelinks]{hyperref}

\usepackage{lineno}

\newcommand\ionn[2]{#1$\;${\scshape{#2}}}
\newcommand{\be}{\begin{equation}}
\newcommand{\ee}{\end{equation}}
\newcommand{\bea}{\begin{eqnarray}}
\newcommand{\eea}{\end{eqnarray}}
\def\kms{\ {\rm km\, s}^{-1}}

\def\msun{M_\odot}

\shortauthors{CONROY ET AL.}
\shorttitle{Birth of the Galactic Disk}

\begin{document}
\begin{CJK*}{UTF8}{gbsn}

%---------------------------------------------------------%

\title{Birth of the Galactic Disk Revealed by the H3 Survey}

\author{\vspace{-1.5cm} Charlie Conroy\altaffilmark{1}, David
  H. Weinberg\altaffilmark{2}, Rohan P.  Naidu\altaffilmark{3,4}, Tobias
  Buck\altaffilmark{5,6}, James W. Johnson\altaffilmark{7}, Phillip
  Cargile\altaffilmark{1}, Ana Bonaca\altaffilmark{7}, Nelson
  Caldwell\altaffilmark{1}, Vedant Chandra\altaffilmark{1}, Jiwon
  Jesse Han\altaffilmark{1}, Benjamin D. Johnson\altaffilmark{1},
  Joshua S. Speagle\altaffilmark{8,9,10,11}, Yuan-Sen
  Ting\altaffilmark{2,12,13}, Turner Woody\altaffilmark{1}, Dennis
  Zaritsky\altaffilmark{14}}

\altaffiltext{1}{Center for Astrophysics $\vert$ Harvard \& Smithsonian, Cambridge, MA, 02138, USA}
\altaffiltext{2}{Department of Astronomy \& CCAPP, The Ohio State University,
  140 W. 18th Avenue Columbus, OH 43210, USA}
\altaffiltext{3}{NASA Hubble Fellow}
\altaffiltext{4}{MIT Kavli Institute for Astrophysics and Space Research, 77 Massachusetts Ave., Cambridge, MA 02139, USA}
\altaffiltext{5}{Interdisziplinares Zentrum fur Wissenschaftliches
  Rechnen, Universitat Heidelberg, Im Neuenheimer Feld 205, D-69120
  Heidelberg, Germany}
\altaffiltext{6}{Zentrum fur Astronomie, Institut fur Theoretische Astrophysik, Universitat Heidelberg, Albert-Ueberle-Strasse 2, D-69120 Heidelberg, Germany}
\altaffiltext{7}{The Observatories of the Carnegie Institution for
  Science, 813 Santa Barbara St., Pasadena, CA 91101, USA}
\altaffiltext{8}{Banting \& Dunlap Fellow}
\altaffiltext{9}{Department of Statistical Sciences, University of Toronto, Toronto, ON M5S 3G3, Canada}
\altaffiltext{10}{David A. Dunlap Department of Astronomy \& Astrophysics, University of Toronto, Toronto, ON M5S 3H4, Canada}
\altaffiltext{11}{Dunlap Institute for Astronomy \& Astrophysics, University of Toronto, Toronto, ON M5S 3H4, Canada}
\altaffiltext{12}{Research School of Astronomy \& Astrophysics,
  Australian National University, Cotter Road, Weston Creek, ACT 2611,
  Canberra, Australia}
\altaffiltext{13}{Research School of Computer Science, Australian
  National University, Acton ACT 2601, Australia}
\altaffiltext{14}{Steward Observatory, University of Arizona, 933 North Cherry
 Avenue, Tucson, AZ 85721, USA}

\begin{abstract}

  We use chemistry ([$\alpha$/Fe] and [Fe/H]), main sequence turnoff
  ages, and kinematics determined from H3 Survey spectroscopy and {\it
    Gaia} astrometry to identify the birth of the Galactic disk.  We
  separate in-situ and accreted stars on the basis of angular momenta
  and eccentricities.  The sequence of high$-\alpha$ in-situ stars
  persists down to at least [Fe/H] $\approx-2.5$ and shows unexpected
  non-monotonic behavior: with increasing metallicity the population
  first declines in [$\alpha$/Fe], then increases over the range
  $-1.3\lesssim$ [Fe/H] $\lesssim-0.7$, and then declines again at
  higher metallicities.  The number of stars in the in-situ population
  rapidly increases above [Fe/H] $\approx-1$.  The average kinematics
  of these stars are hot and independent of metallicity at [Fe/H]
  $\lesssim-1$ and then become increasingly cold and disk-like at
  higher metallicities.  The ages of the in-situ, high$-\alpha$ stars
  are uniformly very old ($\approx13$ Gyr) at [Fe/H] $\lesssim-1.3$,
  and span a wider range ($8-12$ Gyr) at higher metallicities.
  Interpreting the chemistry with a chemical evolution model suggests
  that the non-monotonic behavior is due to a significant increase in
  star formation efficiency, which began $\approx13$ Gyr ago.  These
  results support a picture in which the first $\approx1$ Gyr of the
  Galaxy was characterized by a ``simmering phase'' in which the star
  formation efficiency was low and the kinematics had substantial
  disorder with some net rotation.  The disk then underwent a dramatic
  transformation to a ``boiling phase'', in which the star formation
  efficiency increased substantially, the kinematics became disk-like,
  and the number of stars formed increased tenfold.  We interpret this
  transformation as the birth of the Galactic disk at $z\approx4$.
  The physical origin of this transformation is unclear and does not
  seem to be reproduced in current galaxy formation models.

\end{abstract}

%\keywords{Galaxy: halo --- Galaxy: kinematics and dynamics}

%---------------------------------------------------------%

\section{Introduction}
\label{s:intro}

A thorough understanding of the components of the Galaxy, including
their formation channel(s) and dynamical evolution, is of great
interest for several reasons.  As the best-studied galaxy in the
universe, the Milky Way provides stringent tests of galaxy formation
models \citep[e.g.,][]{Guedes11, Wetzel16, Grand17, Buck21}.  The
kinematics of stars in the Galaxy can be very precisely measured, and
therefore may offer novel constraints on the nature dark matter
\citep[e.g.,][]{Ibata02, Johnston02, Bovy17, Bonaca19}.  There is also
a basic and deep human desire to understand our origins, from the
smallest scales to the Galactic ecosystem.

The rotationally-supported disk contains the majority of the stars in
the Galaxy.  It has long been known that the disk harbors multiple
components (or populations) in the space of chemistry, kinematics,
spatial extent, and age \citep[e.g.,][]{Gilmore83, Norris85, Chiba00,
  Nissen10, Bovy12, Haywood13, Hayden15}; see \citet{Rix13} and
\citet{Bland-Hawthorn16} for recent reviews.  Spatially, there is
evidence for thin and thick disks with different scale heights.
Chemically, a clear dichotomy exists in the abundance of $\alpha$
elements relative to iron ([$\alpha$/Fe]) -- a high$-\alpha$ and a
low$-\alpha$ population.  In general, the former has older ages and a
larger scale-height than the latter.  While the correspondence between
high$-\alpha$ and thick disk and low$-\alpha$ and thin disk is
reasonably strong in the solar vicinity, this correspondence breaks
down especially strongly at larger galactocentric distances
\citep[e.g.,][]{Hayden15}.

The origin of these two disk populations has been the subject of much
debate.  One scenario envisions the chemical dichotomy as the result
of smooth evolution of the Galaxy in terms of its star formation rate
(SFR), gas mass, etc., combined with radial mixing, kinematic heating,
or a decline in the vertical dispersion of the gas with time to
explain the observed correlations between kinematics, chemistry, and
scale-height \citep[e.g.,][]{Schonrich09, Loebman11, Bird13,
  Minchev15, Sharma21}.  A second scenario invokes a violent event
such as a merger as the origin of the dichotomy
\citep[e.g.,][]{Reddy06, Villalobos08, Grand18b, Mackereth18,
  Bonaca20, Buck20, Agertz21}.  In this picture the high$-\alpha$ disk
forms first and is truncated at $z\sim1$ by a merger that heats the
existing disk stars and delivers fresh gas to fuel the formation of
the low$-\alpha$ disk.  A related idea is the two-infall model
\citep{Chiappini97, Spitoni19}, which posits two time-separated phases
of accretion with an intervening lull of star formation.  This model
is agnostic regarding the origin of the two-phase accretion history.
A third scenario suggests that most or all of the thick disk was
formed directly from the accretion of disrupting satellite galaxies on
prograde orbits \citep{Abadi03}.

Thanks to the combination of {\it Gaia} astrometry and large
spectroscopic surveys, our understanding of stars on halo-like orbits
has improved dramatically over the past several years.  One striking
discovery was the identification of stars with chemistry
indistinguishable from the high$-\alpha$ disk with halo-like orbits
\citep[e.g.,][]{Bonaca17, Haywood18, DiMatteo19, Belokurov20,
  Bonaca20}.  One interpretation of this population is that the stars
were born in-situ and subsequently dynamically heated to halo-like
orbits, a phenomenon predicted by some simulations
\citep[e.g.,][]{Zolotov09}.  The age distribution of these ``in-situ
halo'' stars is truncated at lookback times of $8-11$ Gyr
\citep{Gallart19, Bonaca20, Belokurov20, Xiang22}, suggesting that the
heating event, perhaps a merger, occurred at $z\sim1-2$ \citep[see
also][]{Vincenzo19, Chaplin20, Montalban21}.

A large number of accreted systems have also been identified as
comprising the accreted stellar halo \citep[e.g.,][]{Belokurov18,
  Helmi18, Koppelman18, Myeong19a, Naidu20a, Horta21}.  By far the
most significant is the {\it Gaia}-Sausage Enceladus (GSE) system,
which comprises most of the accreted halo within 30 kpc of the
Galactic center.  \citet{Naidu21} combined a suite of $N-$body
simulations and kinematic and age data from the H3 Survey
\citep{Conroy19a} to argue that the GSE merger completed approximately
8 Gyr ago, and gave rise to the in-situ halo by heating pre-existing
disk stars.  There is circumstantial evidence for another major
accretion event in the inner Galaxy, perhaps as massive as GSE
\citep[e.g.,][]{Kruijssen19, Massari19, Horta21, Naidu22a}.  It is
possible that there are not one but several accreted structures in the
inner Galaxy.  In any event, it is likely that the accreted stars in
the inner Galaxy were brought in from a merger that occurred prior to
GSE, i.e., at $z>1$.

While the characterization and origins of the main disk and halo
populations are coming into focus, the very earliest epochs of the
Galaxy and the emergence of the disk are still poorly understood.
Stellar metallicity is often invoked as a proxy for age, and so
studying the most metal poor stars may offer clues to early epochs.
Stars associated with a disk component of our Galaxy have indeed been
found at low metallicities \citep[e.g.,][]{Norris85, Morrison90,
  Chiba00, Carollo19, Naidu20a}.  This population, sometimes known as
the metal-weak thick disk, has been identified at metallicities as low
as [Fe/H] $=-2.5$, and generally has hotter kinematics than the
canonical thick disk at higher metallicity \citep{Carollo19}.
Recently, several authors have identified very metal-poor stars
([Fe/H] $<-2.5$) with low eccentricities reminiscent of the thin disk
\citep{Sestito20, Carter21}.  Using [Al/Fe] from APOGEE
\citep{Majewski17} to separate in-situ from accreted stars,
\citet{Belokurov22} identified in-situ stars to [Fe/H] $\approx-1.5$.
The average tangential velocity of the in-situ stars in their sample
increased rapidly at [Fe/H] $\approx-1$.  They identify this
transition as the epoch when the Galaxy spun up from a relatively
disordered state to well-ordered rotation.

In this paper we use H3 spectroscopy and {\it Gaia} astrometry to
identify and characterize in-situ stars from $-2.5\lesssim$ [Fe/H]
$\lesssim0.5$.  We chart the chemistry in terms of the abundance ratio
[$\alpha$/Fe], and we report main sequence turnoff and subgiant ages
and average kinematic behavior as a function of metallicity.  These
measurements allow us to identify the birth of the Galactic disk
associated with a major transformation in the disk that began
$\approx13$ Gyr ago when the mean metallicity was [Fe/H]
$\approx-1.3$.

\begin{figure*}[!t]
\center
\includegraphics[width=1.0\textwidth]{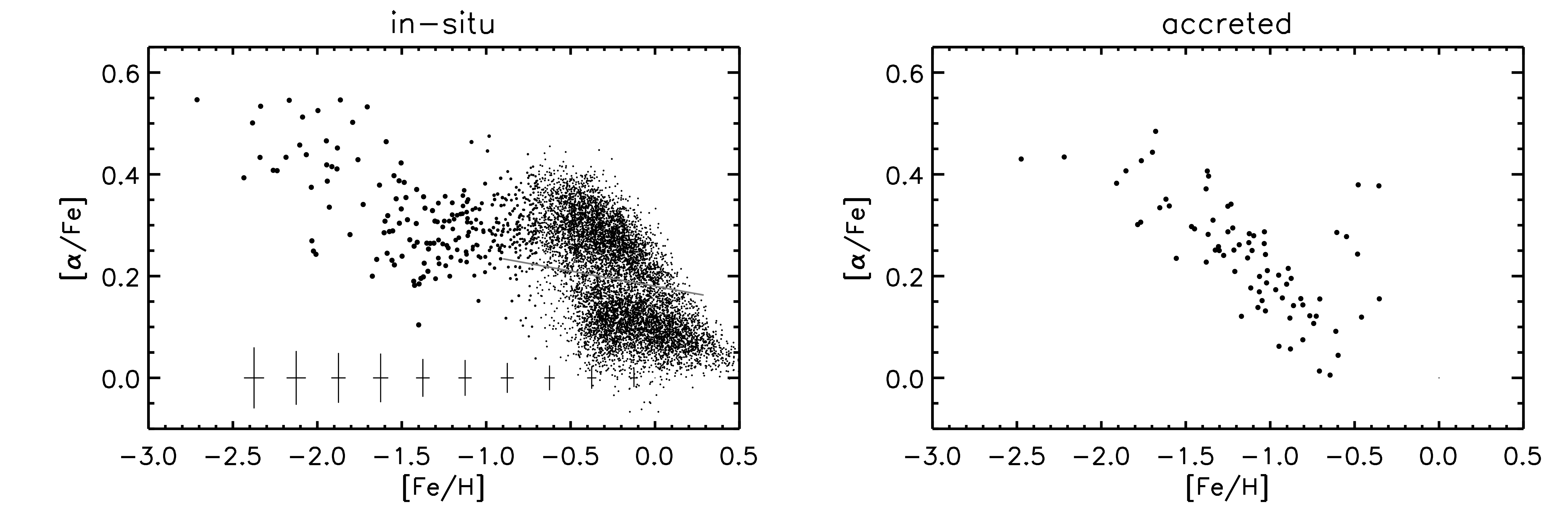}
\vspace{0.1cm}
\caption{Chemistry of in-situ and accreted stars from the H3 Survey.
  In-situ stars are defined to have prograde orbits ($L_Z<0$) with
  eccentricities $e<0.8$, while accreted stars are defined to have
  $e>0.9$ or a combination of $e>0.8$ and $L_Z>500$ km s$^{-1}$
  kpc$^{-1}$ (retrograde orbits).  In the left panel, one clearly sees
  the high$-\alpha$ and low$-\alpha$ sequences at [Fe/H]
  $\gtrsim-0.7$.  Moving to lower metallicities the high$-\alpha$
  population declines in [$\alpha$/Fe] until [Fe/H] $\approx-1.3$, at
  which point [$\alpha$/Fe] increases.  The grey line highlights our
  adopted separation between high$-\alpha$ and low$-\alpha$
  populations.  The accreted population (right panel) displays a
  linear decline in [$\alpha$/Fe] with increasing metallicity.  The
  $\approx7$ outlier stars at [Fe/H] $\gtrsim-0.5$ are likely heated
  stars from the in-situ population.  In the left panel, symbol size
  is inversely proportional to metallicity in order to draw attention
  to the low-metallicity sequence, and the average uncertainties are
  shown as a function of metallicity along the bottom.}
\label{fig:chem}
\end{figure*}

\begin{figure}[!t]
\center
\includegraphics[width=0.47\textwidth]{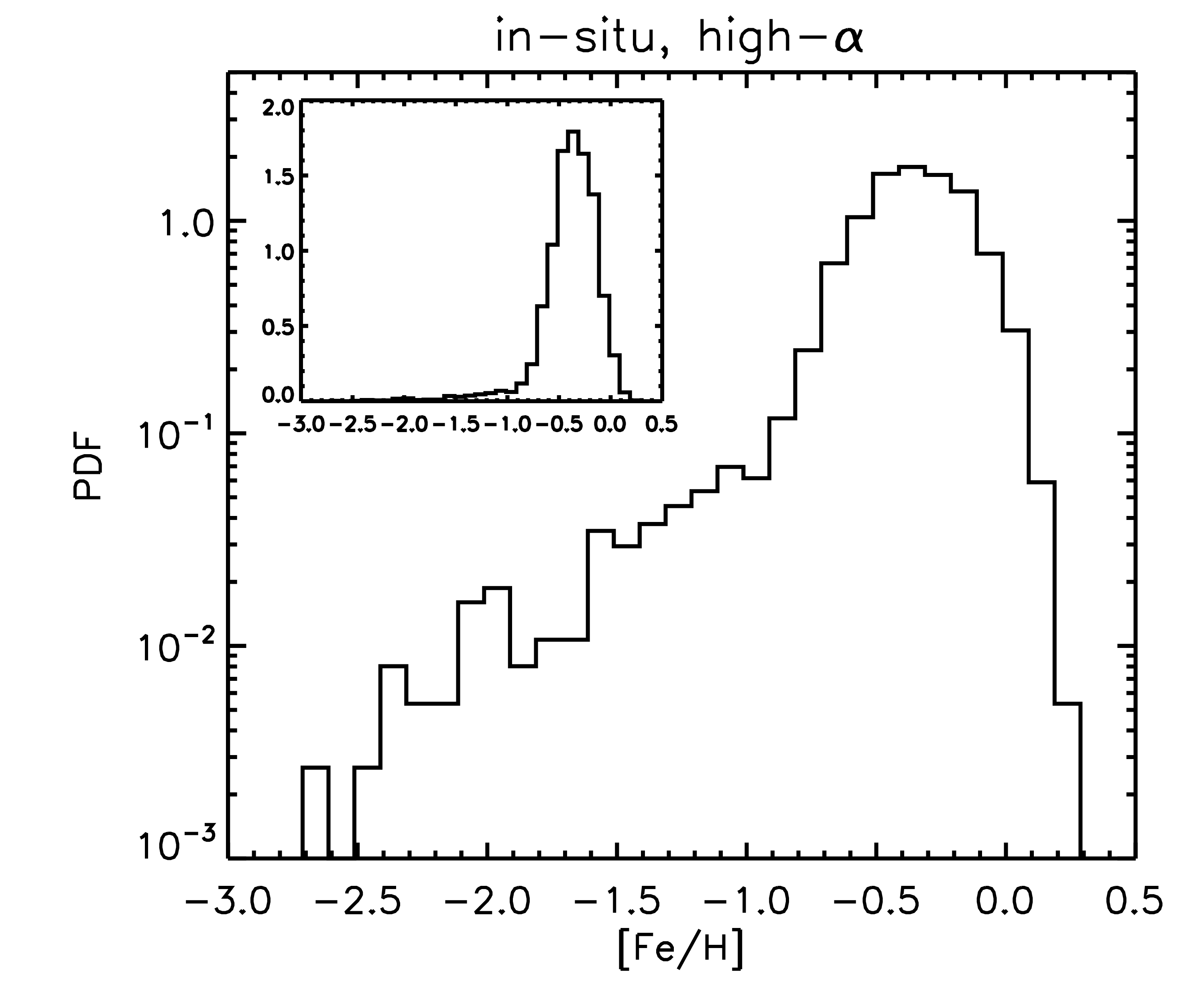}
\vspace{0.1cm}
\caption{Metallicity distribution function (MDF) of the in-situ,
  high$-\alpha$ population.  The inset shows the MDF on a linear
  scale.  There is a substantial increase in stars above a metallicity
  of [Fe/H] $\approx-1.0$.}
\label{fig:mdf}
\end{figure}

\begin{figure*}[!t]
\center
\includegraphics[width=0.95\textwidth]{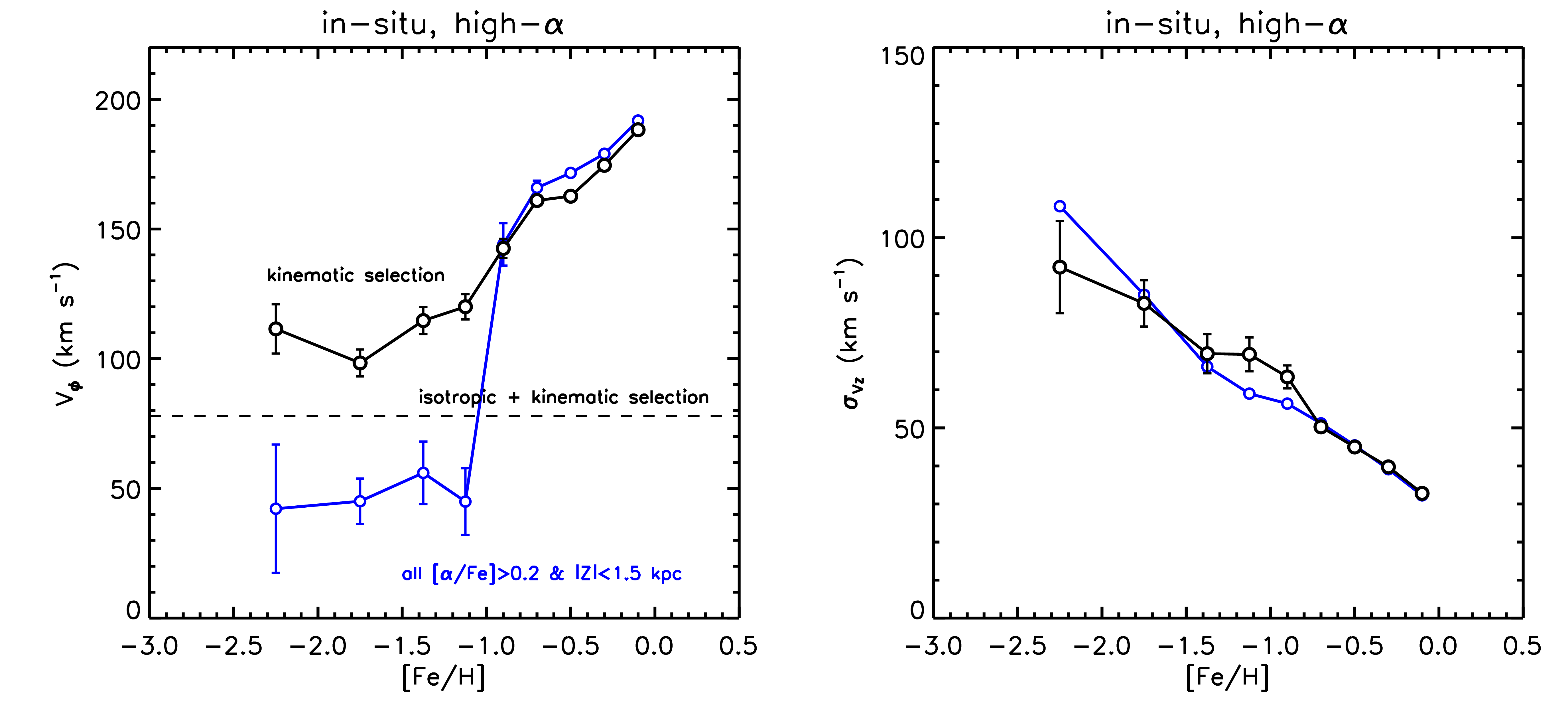}
\vspace{0.1cm}
\caption{Average kinematic behavior of the in-situ, high$-\alpha$
  population as a function of metallicity.  Left panel: average
  azimuthal velocity ($V_\phi$).  Right panel: vertical velocity
  dispersion ($\sigma_{V_z}$).  Our kinematic selection of the in-situ
  sample (black line) is compared to a sample selected only to have
  high $\alpha$ and $|Z|<1.5$ kpc (blue line).  The former selection
  will bias the azimuthal velocity high because stars on retrograde
  and radial orbits are removed, while the latter will bias the
  velocity low because some accreted stars, which have preferentially
  radial orbits, are included.  The true azimuthal velocity of the
  in-situ population lies in between the black and blue lines.  Below
  a metallicity of [Fe/H] $\approx-1$ the population is kinematically
  hot with little net rotation; at higher metallicities the population
  is increasingly cold and disk-like.  We include in the left panel a
  dashed line that indicates the average value for an intrinsically
  isotropic velocity distribution with the in-situ sample selection
  applied.  }
\label{fig:kin}
\end{figure*}

%--------------------------------------------------------%
%--------------------------------------------------------%
%--------------------------------------------------------%

\section{Data}
\label{s:data}

In this paper we use data from the H3 Stellar Spectroscopic Survey
\citep{Conroy19a} and {\it Gaia} EDR3 \citep{GaiaEDR3}.  H3 is
collecting $R=32,000$ spectra over the wavelength range
$5150$\AA$-5300$\AA\, using the Hectochelle spectrograph on the MMT
telescope \citep{Fabricant05, Szentgyorgyi11}.  The survey will
eventually collect 300,000 spectra over $\approx 1500$ uniformly
spaced fields covering $|b|>20^\circ$ and Dec. $>-20^\circ$.  As of
March 2022 the survey has collected 208,000 spectra over 1,100 fields.
The primary survey selection function is based solely on magnitude
($15<r<18$) and {\it Gaia} parallax (a selection that has evolved from
$\pi<0.5$ mas to $\pi<0.3$ mas as the {\it Gaia} data quality has
improved).  The survey includes additional rare high-value targets and
filler targets that are both fainter and at higher parallax than the
main sample.  The high parallax filler sample constitutes
$\approx60$\% of the sample used below.

Stellar parameters are derived using the \texttt{MINESweeper} program
\citep{Cargile20}.  \texttt{MINESweeper} simultaneously fits the
high-resolution spectrum and the broadband photometry (from SDSS,
Pan-STARRS, {\it Gaia}, 2MASS, {\it WISE}) to a library of synthetic
spectral grids and isochrones.  Derived parameters include radial
velocities, distances, reddening, rotational line broadening,
$V_{\rm rot}$, [Fe/H], and [$\alpha$/Fe].  In the model fitting we
assume that O, Ne, Mg, Si, Ar, Ca, and Ti vary in lock-step and we
refer to them as $\alpha$ elements.  The fitted spectral region
contains the \ionn{Mg}{i} triplet as well as many \ionn{Fe}{i} and
\ionn{Fe}{ii} lines, so [$\alpha$/Fe] is primarily determined by Mg
and Fe lines.  \texttt{MINESweeper} employs a Bayesian framework and
includes the {\it Gaia} parallax as a prior on the distance.  The
standard pipeline includes a complex prior on the stellar age
\citep[see][for details]{Cargile20}.  Here, since we are particularly
interested in ages, we use an alternative version of the pipeline in
which the age prior is flat from $4-14$ Gyr.  We have also tested the
effect of the isochrones on the abundances by fitting the spectra to a
version of \texttt{MINESweeper} in which the spectroscopic parameters
are determined without the requirement that the star reside on an
isochrone.  For the sample discussed below, the differences are small
and within the reported uncertainties ($\approx 0.01-0.02$) for [Fe/H]
and [$\alpha$/Fe].

The \texttt{MIST} isochrones \citep{Choi16}, which we use in this
work, include the effects of diffusion on the surface abundances
(specifically, we use \texttt{MIST} v2.1 isochrones, which include
several improvements and extension to [$\alpha$/Fe] variation; see
Dotter et al. in prep for details).  This effect is largest around the
main sequence turnoff \citep{Dotter17}, where much of our sample is
located.  \texttt{MINESweeper} adopts the initial metallicity and
composition as the free variables, but the surface abundance is used
when comparing to the data.  The program therefore returns both the
surface and the initial abundances.  The latter are used in this work
as they are more fundamental and should exhibit less variation than
surface abundances.

The spectrophotometric distances, radial velocities, and {\it Gaia}
proper motions are used to derive a variety of quantities including
projections of the angular momentum vector onto the Galactocentric
coordinate system and the azimuthal velocity, $V_\phi$, in spherical
coordinates.  We adopt the Galactocentric frame implemented in
\texttt{Astropy v4.0} \citep{astropy1, astropy2}.  This frame is
right-handed, i.e., prograde (retrograde) orbits have $L_{\rm{Z}}<0$
($L_{\rm{Z}}>0$).  Orbit-related quantities including eccentricities,
$e$, are computed using \texttt{gala v1.1} \citep{gala1, gala2} with
its default \texttt{MilkyWayPotential}.  See \citet{Naidu20a} for
details.

In this paper we focus on a high-quality subset of the H3 data.  In
particular, we require spectroscopic SNR $>20$, log $g>3.5$,
$V_{\rm rot}<2 \kms$, {\it Gaia} RUWE $<1.5$, and
$\chi^2_{\rm spec}/{\rm DOF}<2.5$.  There are a variety of data
quality flags; we require that no flags have been set.  The SNR, flag,
and log $g$ cuts reduce the sample to 9,476 stars.  The RUWE cut
reduces the sample to 9,163 stars, and the last two cuts result in a
final sample of 8,544 stars.  For this sample, the median SNR of the
{\it Gaia} EDR3 parallax and proper motion is $17$ and $>100$,
respectively.  The median formal uncertainties on [Fe/H] and
[$\alpha$/Fe] are 0.02, although they increase toward lower
metallicities such that the median uncertainties are $0.05$ at [Fe/H]
$=-2$.

Owing to the parallax selection of the overall survey combined with
the subsequent log $g$ and SNR selections, the distribution of $|Z|$
for this sub-sample peaks at 1.3 kpc, drops sharply at $|Z|<0.7$ kpc,
and has a tail to $|Z|\approx4$ kpc.  The sample is thus ideally
suited to study the high$-\alpha$ disk, which has a scale-height of
$0.6-1$ kpc \citep[e.g.,][]{Bovy12}.

%--------------------------------------------------------%
%--------------------------------------------------------%
%--------------------------------------------------------%

\section{Results}
\label{s:res}

\subsection{Chemistry, kinematics, and ages of in-situ stars}
\label{s:mainres}

We begin by separating stars into in-situ and accreted populations.
These two populations are shown in Figure \ref{fig:chem} in
[$\alpha$/Fe] vs. [Fe/H].  The separation is based solely on kinematic
quantities, but the choice of cuts was determined by inspecting the
distribution of stars in this space.  In particular, the definitions
were chosen such that high eccentricity stars at [$\alpha$/Fe]$<0.2$
and [Fe/H] $\lesssim-0.5$ only appear in the accreted population.  The
accreted sample is defined by the following criteria:
\noindent
\be
e>0.9\, \vee \,  (e>0.8 \, \wedge \, L_Z>500 \,{\rm  km\, s}^{-1}\,{\rm  kpc}^{-1}),
\ee
where $e$ is the orbital eccentricity and $L_Z$ is the $Z-$component
of the angular momentum vector (recall that $L_Z>0$ means
retrograde orbits in our convention).  The in-situ population is
defined by:
\be
e<0.8\, \wedge\, L_Z < 0\, {\rm km \, s}^{-1}\, {\rm kpc}^{-1}.
\ee
\noindent
We also remove one star that has an orbit apocenter $r_{\rm apo}>30$
kpc.  The union of these two selections does not cover all of
parameter space - there is a transition region containing an overlap
of accreted and in-situ stars (7\% of the sample).  The $\approx7$
metal-rich stars in the accreted sample at [Fe/H] $\gtrsim -0.6$ are
likely heated stars from the in-situ population (they mostly overlap
with the metal-rich, high$-\alpha$ in-situ population).  For reasons
discussed below, it is likely that the in-situ sampled defined here is
incomplete but relatively pure.  Such a sample is well-suited to study
abundance patterns and ages.

In principle there could be accreted stars on prograde orbits that
would contaminate our in-situ selection.  To test this possibility, we
have divided our in-situ sample into low and moderate-eccentricity
subsamples ($e<0.4$ and $e>0.6$).  These two sub-populations trace the
same sequence in abundance space.  It is unlikely that an accreted
population would have the same distribution in abundance-eccentricity
space as the in-situ population, so we conclude from this test that
contamination of the in-situ population by accreted stars is low.

Figure \ref{fig:chem} demonstrates that in-situ (prograde) stars span
the full range of metallicities probed by H3.  Importantly, the
sequence of stars at high $\alpha$ smoothly extends from low
metallicities to the traditional high$-\alpha$ locus at [Fe/H] $>-0.5$
and [$\alpha$/Fe]$>0.2$.  In detail, the chemistry of the
high$-\alpha$ sequence is complex and shows three distinct regimes:
declining [$\alpha$/Fe] with increasing metallicity at [Fe/H]
$\lesssim-1.3$; rising [$\alpha$/Fe] to [Fe/H] $\approx-0.7$, and
declining [$\alpha$/Fe] again for [Fe/H]$\gtrsim-0.7$.  This
non-monotonic chemical track is the primary observational result of
this paper.

In the rest of this paper we focus on the high$-\alpha$ subset of the
in-situ population.  We define the high$-\alpha$ sequence as
[$\alpha$/Fe] $>-0.06$ [Fe/H] $+0.18$ for [Fe/H] $>-0.9$ (indicated by
the grey line in the left panel of Figure \ref{fig:chem}) and all
in-situ stars at [Fe/H] $\le-0.9$.  The results below are unchanged if
we adopt a simpler selection of [$\alpha$/Fe] $>0.2$ for the
high$-\alpha$ population.

Figure \ref{fig:mdf} shows the metallicity distribution function (MDF)
of the in-situ high$-\alpha$ stars.  The H3 selection function (being
a function of parallax and magnitude) is nearly unbiased with respect
to metallicity.  We therefore interpret the observed MDF as a fair
representation of the true underlying MDF in the volume probed by the
sample.  One clearly sees a rapid increase in the number of stars in
this population at [Fe/H] $\gtrsim-1.0$.  Only 4\% of stars in the
in-situ high$-\alpha$ population have [Fe/H] $<-1.0$ and only 1\% have
[Fe/H] $<-1.3$, \citep[in good agreement with][who find 1\% of in-situ
stars have [Fe/H $<-1$]{Belokurov22}.

Figure \ref{fig:kin} shows the average kinematic behavior of the
in-situ high$-\alpha$ population as a function of metallicity.  In
this figure we have relaxed the SNR threshold from 20 to 15, which
increases the sample from 8,544 to 17,565 stars.  The kinematic
uncertainties remain small at lower SNR, allowing us to increase the
sample size here.  The left panel shows the average azimuthal velocity
($V_\phi$) versus metallicity while the right panel shows the vertical
velocity dispersion versus metallicity.  The left panel displays two
clear regimes: at [Fe/H] $\lesssim-1$ the kinematic behavior is
constant, while at [Fe/H] $\gtrsim-1$ the population becomes
increasingly well-ordered (and kinematically colder) toward higher
metallicities.  This change in behavior of the kinematics at [Fe/H]
$\approx-1$ is in good agreement with \citet{Belokurov22}.  The trend
at [Fe/H] $\gtrsim-1$ is broadly in agreement with previous work
\citep[e.g.,][]{Lee11, ReFiorentin19, Han20}.

\begin{figure*}[!t]
\center
\includegraphics[width=0.95\textwidth]{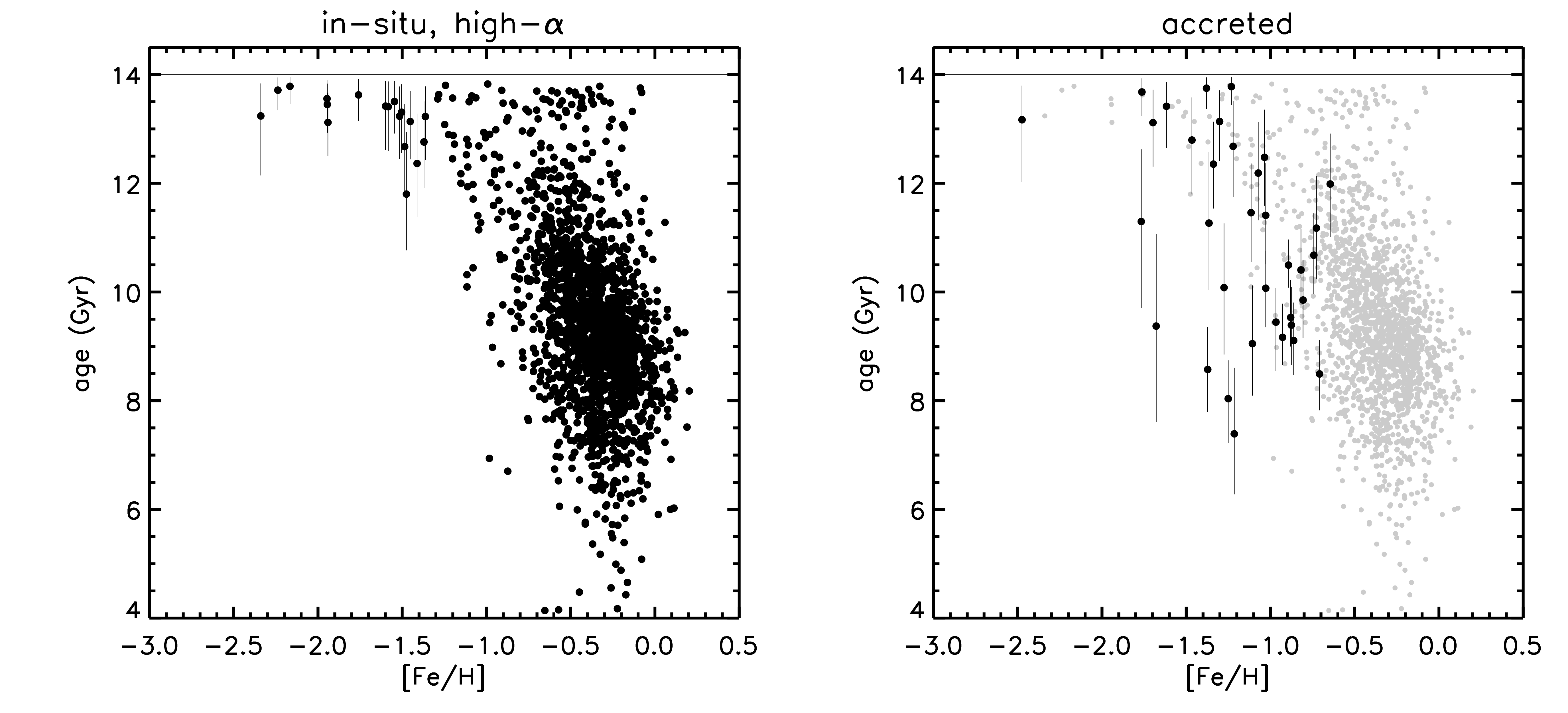}
\vspace{0.1cm}
\caption{Ages of the in-situ, high$-\alpha$ (left panel) and accreted
  (right panel) stars as a function of metallicity.  The sample here
  is restricted to main sequence turnoff and subgiant stars
  ($3.8<{\rm log}\ g<4.2$) where age estimates are most reliable.  The
  upper limit on allowed ages is 14 Gyr (solid line).  Error bars
  represent formal 68\% confidence intervals; in the left panel they
  are shown only at [Fe/H] $<-1.3$ for clarity.  In the right panel,
  the in-situ stars are shown as light grey points for direct
  comparison.  The in-situ, high$-\alpha$ population is uniformly very
  old at [Fe/H] $\lesssim-1.3$, with ages $\gtrsim 12$ Gyr.  At
  metallicities above [Fe/H] $\approx-1$ the ages span a wide range,
  from $\approx8-12$ Gyr.  In contrast to the in-situ population, the
  accreted population contains stars younger than 12 Gyr even at low
  metallicities.}
\label{fig:ages}
\end{figure*}

Our in-situ population is defined to be prograde ($L_Z<0$) and have
$e<0.8$, so care must be taken when interpreting the kinematics of
this population.  To provide some guidance, we have performed two
tests.  First, in Figure \ref{fig:kin} we compare the standard
kinematic selection of the in-situ stars (black lines) to a selection
based solely on chemistry ([$\alpha$/Fe] $>0.2$) and height from the
Galactic plane ($|Z|<1.5$ kpc; blue lines).  We can see from Figure
\ref{fig:chem} that the chemistry cut retains nearly all of the
in-situ high$-\alpha$ sequence, while removing the metal-rich end of
the accreted population.  The cut on $Z$ is designed to remove
additional accreted stars, as they extend to larger $|Z|$ than the
kinematically-selected in-situ stars.  We expect the kinematic
selection to bias the resulting average velocities high, as in-situ
stars on retrograde or radial orbits are removed, while the chemistry
and spatial selection will bias the velocities low, as there will be
contamination from accreted stars, which preferentially have radial
orbits.  The true average velocity of the in-situ stars should lie in
between these two limits.  The key point for this paper is that the
azimuthal velocity changes abruptly at [Fe/H] $\approx-1$.

As a second test, we have taken the \citet{Rybizki18} mock stellar
catalog and selected stars with log $g>3.5$ and applied our in-situ
selection ($e>0.8$ and $L_Z<0$) as well as the H3 window function.  We
then computed the average azimuthal velocity for the halo component of
this mock catalog, which intrinsically has no net rotation and a
nearly isotropic velocity ellipsoid.  This is shown as a dashed line
in the left panel of Figure \ref{fig:kin}.  Even at the lowest
metallicities the fiducial kinematic selection (black line) has
average velocities in excess of the null expectation.  For this test
to be informative, the underlying in-situ velocity ellipsoid would
need to be similar to that assumed in the mock catalog.
\citet{Belokurov22} measure dispersions from their low-metallicity
in-situ sample of $\sigma\approx80-90\kms$ in the $R$, $Z$, and $\phi$
components, while the \citet{Rybizki18} model assumes
$\sigma_R=141\kms$, $\sigma_Z=75\kms$, and $\sigma_\phi=75\kms$.
While not identical, these values are quite close, especially in $Z$
and $\phi$, and so we conclude that this test is reliable.  From these
two tests we infer that there is evidence for modest net rotation of
the low-metallicity in-situ population with $V/\sigma\lesssim 1$.

In Figure \ref{fig:ages} we show the age-metallicity relation for the
in-situ high$-\alpha$ (left panel) and accreted (right panel)
populations.  For this figure we restrict the sample to stars with
$3.8<{\rm log}\ g<4.2$ (i.e., the main sequence turnoff and subgiant
branch), where the age constraints are most precise.  The median
formal uncertainty on log age in this regime is 0.03.  Assumptions in
the isochrones are the dominant source of systematic uncertainty,
which we do not attempt to quantify here.  However, we emphasize that
any change to the underlying stellar models is more likely to produce
a produce an overall shift in ages at a given metallicity, rather than
an increase in the scatter.  For the accreted population we also
remove the seven most metal-rich stars that are likely contamination
from the in-situ population and are visible in the right panel of
Figure \ref{fig:chem}.

Figure \ref{fig:ages} shows that the in-situ population is remarkably
old at [Fe/H] $\lesssim-1.3$, with ages almost exclusively $>12$ Gyr,
and a median age of 13.3 Gyr (implying a formation epoch of $z>4$).
Moreover, the range in ages at low metallicity is very small - the
dispersion is 0.5 Gyr, which is comparable to the average uncertainty.
In other words, the data are consistent with the low-metallicity
in-situ stars having formed over a very short interval 13.3 Gyr ago.
This is in stark contrast to the accreted population, which has a much
broader age distribution even at low metallicities.  This comparison
shows that metal-poor stars are not always maximally old and that
\texttt{MINESweeper} is able to distinguish between old ($10-12$ Gyr)
and very old ($>12$ Gyr) ages.  At metallicities above [Fe/H]
$\approx-1$ the ages of the high$-\alpha$ in-situ population span a
much wider range, from $\approx8-12$ Gyr.

There is an interesting population of stars with ages $\gtrsim13$ Gyr
and [Fe/H] $>-1$ that is offset from the main locus at $8-12$ Gyr.
These old, metal-rich stars appear in other samples as well
\citep[e.g.,][]{Edvardsson93, Bensby14}.  We have inspected the
\texttt{MINESweeper} fits of these stars and see no obvious problems.
We speculate that unresolved binaries could move stars slightly
brighter and redder, thereby producing spuriously old ages around the
main sequence turnoff.  The {\it Gaia} RUWE parameter has been
previously used to identify unresolved binaries \citep{Belokurov20b}.
Our parent sample already has a RUWE $<1.5$ cut applied.  We note that
this population of ancient metal-rich stars has a higher mean RUWE
than the overall sample (1.07 vs. 1.02), and plotting only stars with
RUWE $<1.02$ removes a large fraction of this population from Figure
\ref{fig:ages}.  Further inspection of this sample is warranted.

\begin{figure*}[!t]
\center
\includegraphics[width=1.0\textwidth]{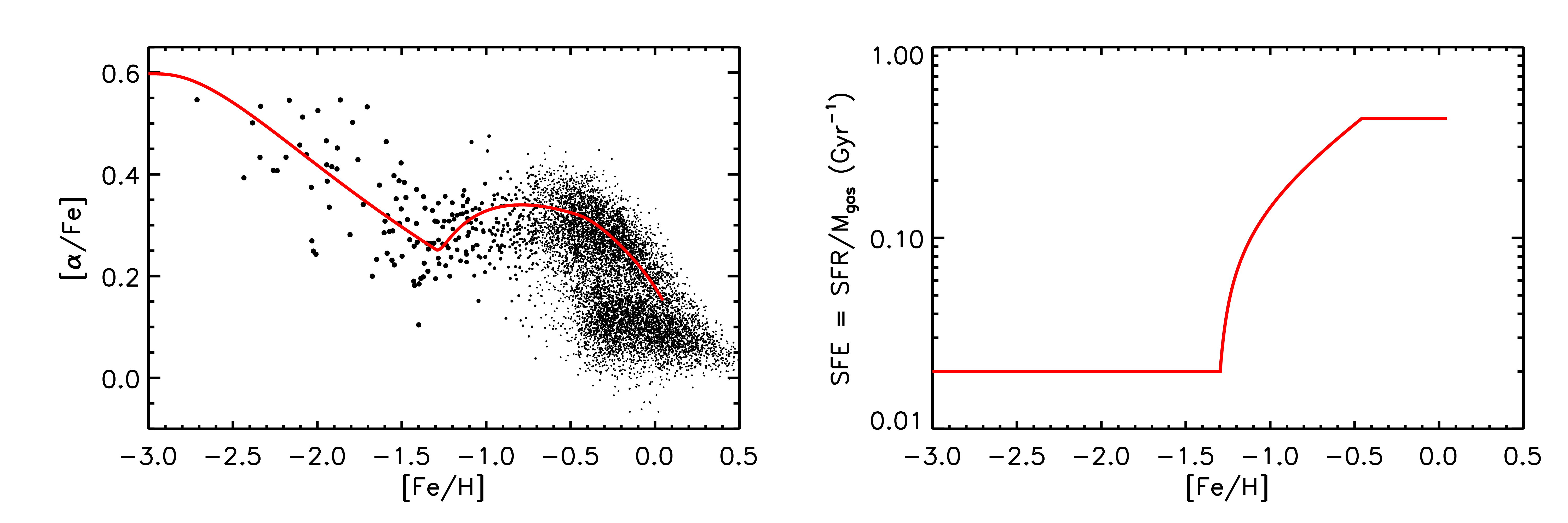}
\vspace{0.1cm}
\caption{Chemical evolution modeling of the in-situ, high$-\alpha$
  population.  The red line shows a model in which the SFR is related
  to the gas mass by the star formation efficiency, whose behavior is
  shown in the right panel.  The model has a constant gas inflow rate
  and a mass-loading factor $\eta\equiv\dot{M}_{\rm out}/{\rm SFR}=2$.
  The key point is that the change in chemical evolution track around
  [Fe/H] $\approx-1.3$ is driven by a large increase in the star
  formation efficiency.}
\label{fig:model}
\end{figure*}

\begin{figure}[!h]
\center
\includegraphics[width=0.47\textwidth]{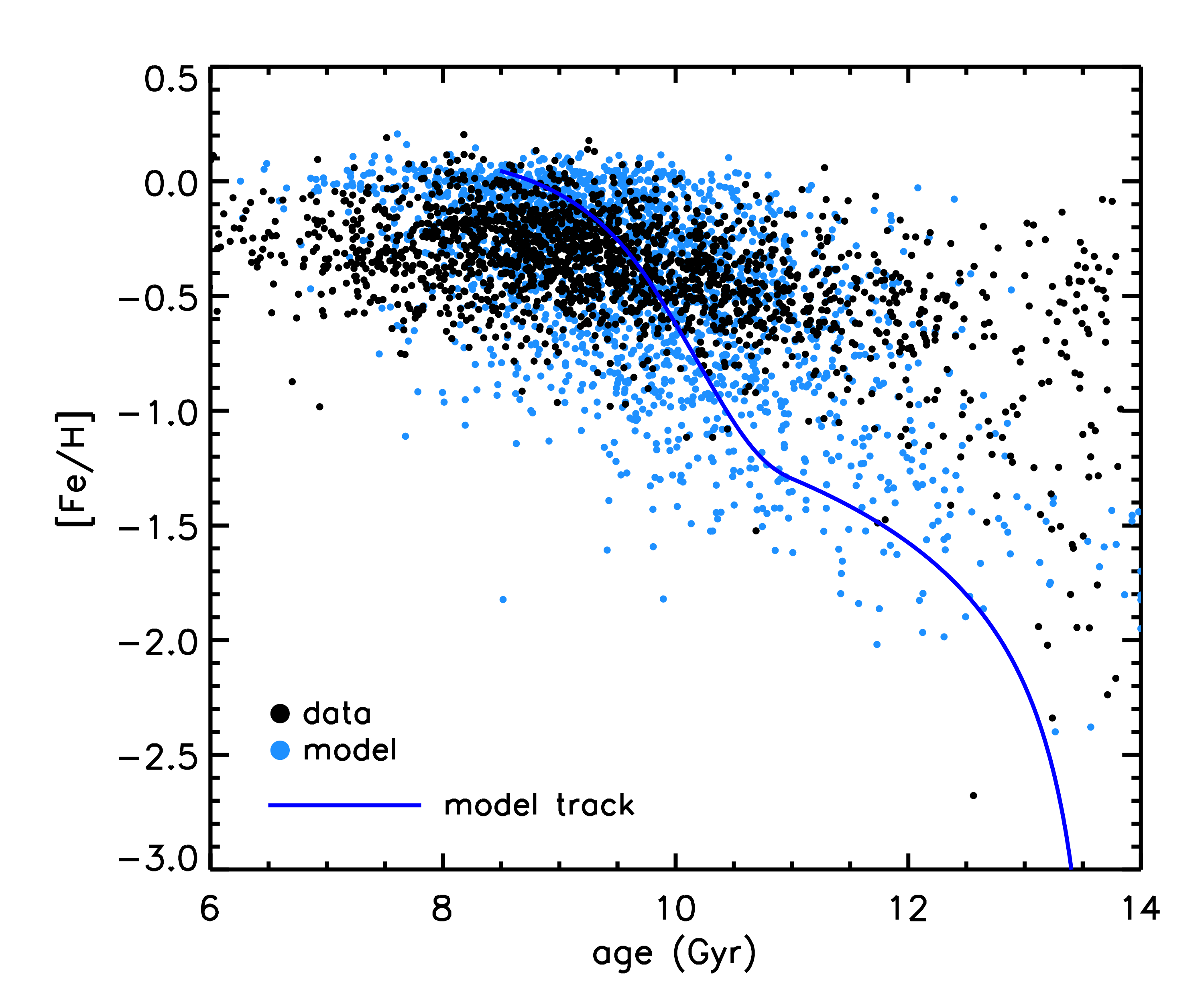}
\vspace{0.1cm}
\caption{Age-metallicity relation comparing the in-situ, high$-\alpha$
data and the chemical evolution model.  In addition to the (noiseless)
model track, we have sampled the model according to the star formation
history and applied measurement uncertainties in both age and
metallicity to mimic the data.  The model produces stars $\approx1$
Gyr younger than the data at [Fe/H]$<-1.3$ but is otherwise in
reasonable agreement with the data.}
\label{fig:model2}
\end{figure}

\vspace{1cm}

\subsection{A Chemical Evolution Model}
\label{s:chemevol}

In this section we employ a chemical evolution model to interpret the
abundance pattern of the in-situ high$-\alpha$ population.  We use the
\texttt{VICE} program \citep{Johnson20}, which is a flexible package
for computing chemical evolution models for a wide range of input
physics (inflows, outflows, varying star formation efficiencies and
yields, etc.).  In particular, we are motivated by \citet{Weinberg17}
to consider changes in the star formation efficiency timescale,
$\tau_\ast$, defined as the gas mass divided by the star formation
rate: $\tau_\ast\equiv M_{\rm gas}/$SFR.  Abrupt changes in
$\tau_\ast$ can result in non-monotonic behavior in [$\alpha$/Fe]
vs. [Fe/H] \citep[e.g.,][]{Weinberg17, Johnson20}, qualitatively
similar to what we observe in the data (Figure \ref{fig:chem}).  Below
we refer to the star formation efficiency (SFE
$\equiv \tau_\ast^{-1}$), which is simply the inverse of $\tau_\ast$.

\begin{figure*}[!t]
\center
\includegraphics[width=1.0\textwidth]{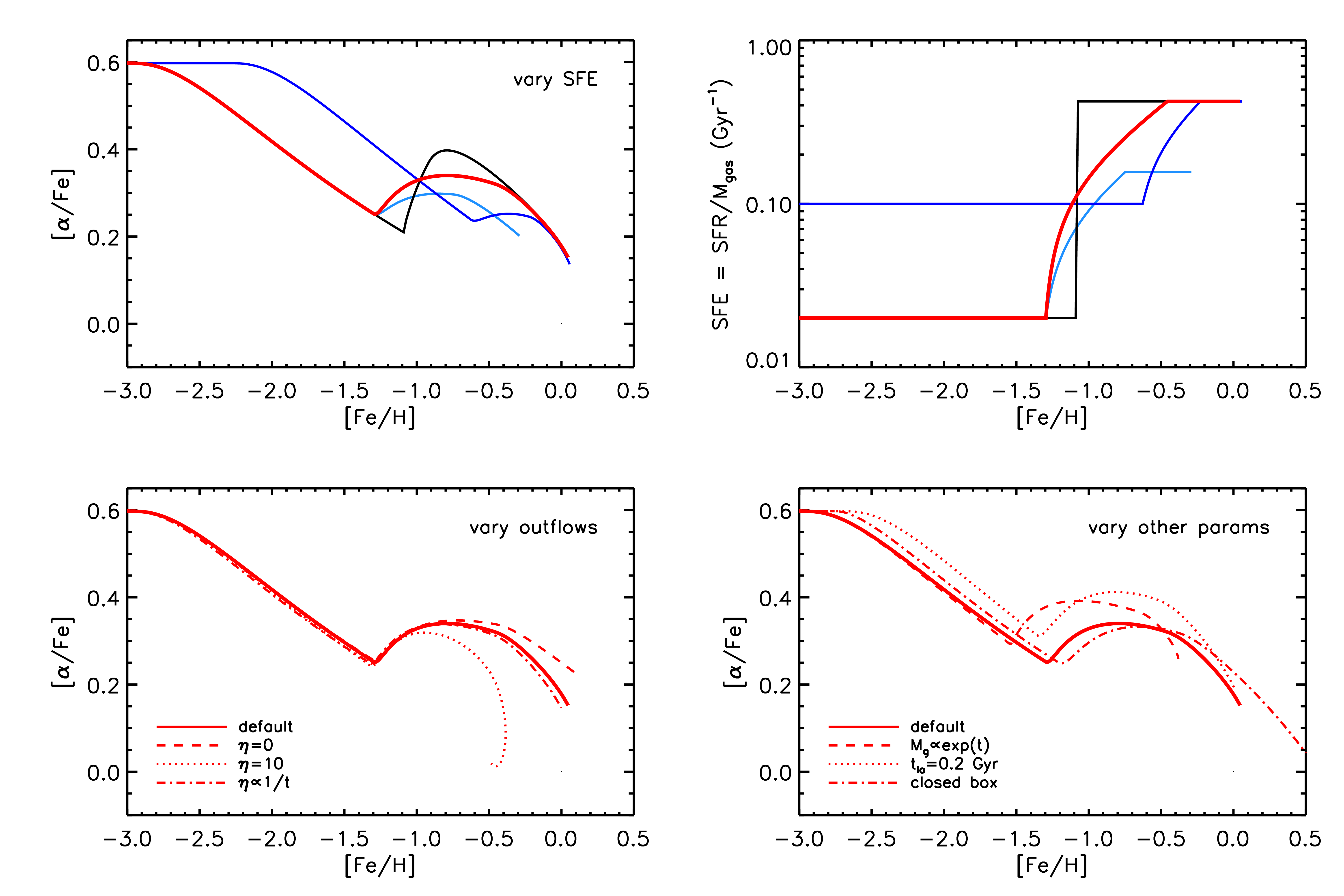}
\vspace{0.1cm}
\caption{Variations to the default chemical evolution model.  The top
  left panel shows the effect of varying the SFE on the abundance
  tracks; the corresponding SFEs are shown in the top right panel.
  The bottom left panel shows variations in the mass-loading factor,
  $\eta$, where the default is $\eta=2$.  The variations in $\eta$ are
  large and include a time-dependent mass-loading factor.  The bottom
  right panel shows changes in the gas inflow rate ($M_g\propto e^t$),
  an increase in the minimum SNe Ia time (from 0.1 to 0.2 Gyr), and
  the adoption of a closed box model (all gas is present at $t=0$ with
  no inflows or outflows).  }
\label{fig:modelvar}
\end{figure*}

The dimensionality and uncertainty in chemical evolution models is
large and includes yields from core-collapse (CC) supernovae (SNe) and
Type Ia SNe, the delay time distribution of Type Ia SNe, parameterized
functions for the gas inflow and outflow rates and SFRs, and the
metallicity of the outflows and inflows.  Motivated by
\citet{Johnson20}, we choose to parameterize the model in terms of the
SFE timescale and evolution of the gas mass with time.  The SFR is
inferred from these two quantities.  The model starts with no gas and
has a constant inflow rate of zero metallicity gas:
$\dot{M}_{\rm gas}=5\,\msun\, {\rm yr}^{-1}$.  The normalization of
the inflow rate has no effect on the tracks in chemistry space and
only affects the amplitude of the SFR.

We adopt the following IMF-averaged Mg and Fe yields for CC and Ia
SNe: $y^{\rm CC}_{\rm Mg}=0.0026$, $y^{\rm CC}_{\rm Fe}=0.0012$,
$y^{\rm Ia}_{\rm Mg}=0.0$, $y^{\rm Ia}_{\rm Fe}=0.003$.  These yields
follow \citet{Johnson20}, except that we increase the Type Ia Fe yield
to better reproduce the slope of [Mg/Fe] vs. [Fe/H] at low metallicity
(Johnson et al. adopt $y^{\rm Ia}_{\rm Fe}=0.0017$).  We adopt
IMF-averaged yields rather than individual star yields because we do
not expect star formation and mixing to be so well-organized in the
chaotic early Galaxy.  We therefore assume that IMF sampling effects
will create scatter, rather than a systematic trend, in the abundance
pattern.  CC yields are uncertain because of effects in massive star
evolution, explosive nucleosynthesis, and which stars collapse to form
black holes rather than exploding (see discussion in
\citealt{Griffith2021}).  SNIa Fe yields are uncertain mainly because
of uncertainties in the normalization of the SNIa rate (see
\citealt{Maoz2017}).  Abundance ratios depend primarily on yield
ratios, so here we are adjusting yields within plausible ranges to
match the observed abundance ratio trends.  Changing all yields by the
same multiplicative factor is largely degenerate with changing the
outflow efficiency \citep{Weinberg17}.  Our most important assumption
is that the IMF-averaged yield ratios are metallicity independent in
the regime of interest, so that the trends of [$\alpha$/Fe] vs. [Fe/H]
seen in Figure \ref{fig:chem} reflect evolution in the ratio of CC to
Ia enrichment, not changes in yields.  We return to this point in
Section \ref{s:caveats}.

We consider a model that has outflows with a mass-loading factor
$\eta\equiv\dot{M}_{\rm out}/$SFR $=2$ where $\dot{M}_{\rm out}$ is the
mass outflow rate.  The outflows are assumed to have the same
metallicity as the star-forming gas.  If we assumed that outflows
preferentially ejected CC products, we would require lower
$y^{\rm Ia}_{\rm Fe}$ to obtain similar [$\alpha$/Fe]-[Fe/H]
evolutionary tracks.  In our model, we interpret the downward trend in
[$\alpha$/Fe] between [Fe/H] $=-2.5$ and $-1.3$ as a consequence of
increasing SNIa enrichment.  Because the timescale for this enrichment
is long -- for a standard delay time distribution, a stellar
population produces half of its Type Ia SNe in about 1 Gyr -- the SFE
during this phase must be very low so that the population does not
evolve past [Fe/H] $=-1.3$.  We interpret the rise in [$\alpha$/Fe]
between [Fe/H] $=-1.3$ and $-0.7$ as a consequence of accelerating star
formation, which increases the rate of CC enrichment relative to Ia
enrichment.  To achieve a good fit to the observed trend, we adopt the
following functional form for $\tau_\ast$ (in units of Gyr):
\noindent
\begin{equation}
  \label{eqn}
  \tau_\ast =
    \begin{cases}
      50 &  t< 2.5 \, {\rm Gyr} \\
      50/[1+3(t-2.5)]^2 &  2.5\le t\le 3.7\, {\rm Gyr} \\
      2.36 &  t> 3.7 \, {\rm Gyr}.
   \end{cases}       
\end{equation}
\noindent
An instantaneous decrease in $\tau_\ast$ at 2.5 Gyr would result in a
rapid increase in [Mg/Fe] at nearly fixed [Fe/H] \citep[see below,
and][]{Weinberg17, Johnson20}.  In order to match the observations we
required a more gradual change; the precise function was determined by
trial-and-error.

\begin{figure*}[!t]
\center
\includegraphics[width=1.0\textwidth]{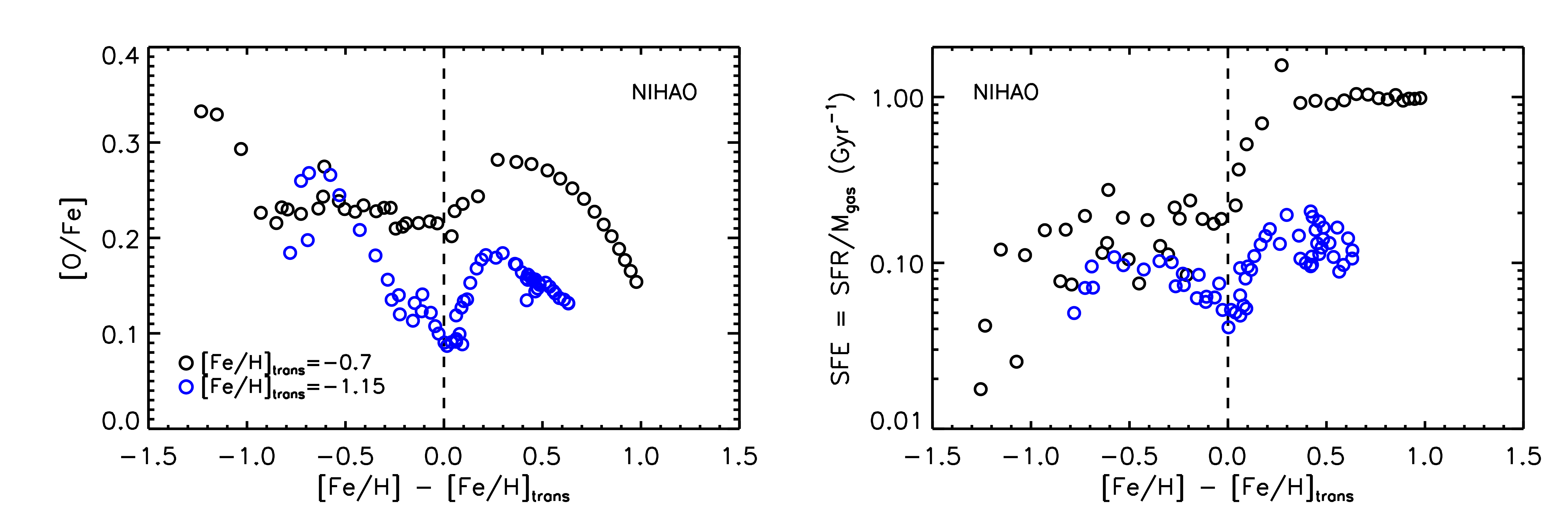}
\vspace{0.1cm}
\caption{Comparison to simulations from the NIHAO project.  Left
  panel: [O/Fe] vs. metallicity for two galaxies that show
  non-monotonic abundance behavior (\texttt{g7.66e11} in black and
  \texttt{g3.59e11} in blue).  Right panel: star formation efficiency
  (SFE) vs. metallicity for these two galaxies.  In each panel, the
  metallicity is shown relative to the metallicity at which the
  non-monotonic transition occurs (labeled [Fe/H]$_{\rm trans}$).  The
  sudden rise in [O/Fe] at the transition scale is clearly associated
  with a rise in SFE in both simulated galaxies.  This provides
  additional support for our interpretation of the observed
  non-monotonic abundance trends in terms of an increase in SFE.}
\label{fig:nihao}
\end{figure*}

Figure \ref{fig:model} shows the resulting model track in
[$\alpha$/Fe] vs. [Fe/H] in comparison to the data.  We adopt [Mg/Fe]
from the model to compare with [$\alpha$/Fe] as the H3 spectral window
is mostly sensitive to Mg.  The model parameters were tuned to fit the
data shown in the left panel, so the agreement there is by
construction.  The right panel shows the model star formation
efficiency, $\tau_\ast^{-1} \equiv {\rm SFR} /M_{\rm gas}$
(Eq. \ref{eqn}) as a function of metallicity.  In this model the CC
yields produce a plateau in [Mg/Fe] of $+0.6$, which is barely visible
in Figure \ref{fig:model} at [Fe/H] $\approx-3$.  The second plateau
at [Mg/Fe] $\approx0.3$ at higher metallicity is a consequence of the
accelerated star formation rate over this interval.

The key conclusion from this model is that the non-monotonic track in
the observed [$\alpha$/Fe] vs. [Fe/H] plane can be explained by a
dramatic and relatively rapid -- though not instantaneous -- increase
in the star formation efficiency.  Typical values for $\tau_\ast$ in
the local universe are $\approx2$ Gyr for the molecular gas
\citep{Leroy08}, comparable to the asymptotic value of $\tau_\ast$ in
our model.  Much higher values of $\tau_\ast$ could arise if most of
the cold gas is in atomic form \citep[see discussion in][]{Johnson20}.
Alternatively, a low effective $\tau_\ast$ could arise if metals from
supernovae are ejected into the gaseous halo and thus diluted by a gas
reservoir that is much larger than the reservoir that is actually
forming stars \citep{Mason24}.

A similar non-monotonic track is apparent in the two-infall+outflow
chemical evolution model of \citet{Brusadin13}.  In their model the
non-monotonic behavior is the result in a significant period ($1-2$
Gyr) in which there is no star formation.  \citet{Micali13} also
report non-monotonic behavior in their three-infall chemical evolution
model, in which the halo, thick disk, and thin disk components form
via three distinct infall episodes.  In their model the non-monotonic
behavior is explicitly linked to a rapid increase in star formation
efficiency.  Alternatively, motivated partly by the preprint version
of our paper, \cite{Chen24} present a scenario in which the
non-monotonic evolution of [$\alpha$/Fe] is triggered by a rapid
acceleration of the gas inflow rate.

Figure \ref{fig:model2} shows the age-metallicity relation comparing
the data and chemical evolution model.  The mock track is shown as a
solid line, and we also include a realization of the model by drawing
model stars proportional to the star formation history and then
applying observational uncertainties.  There is reasonable agreement at
[Fe/H]$>-1.3$ but at lower metallicities the model is approximately 1
Gyr too young.  The model was not fit to the ages, so the poor
agreement is perhaps not surprising.  

Figure \ref{fig:modelvar} shows the effect of various model parameters
on the model [$\alpha$/Fe] vs. [Fe/H] track.  The top panels show the
effect of varying the SFE (left panel is the abundance plane, right
panel is the SFE).  As expected, an instantaneous change in SFE
results in a sharp upward transition in abundance space.  Models with
larger SFE at early times result in a longer plateau phase at low
metallicity and a much less pronounced non-monotonic feature at high
metallicity.  Models with a lower SFE at late times also produce a
less pronounced non-monotonic feature.

In the lower left panel of Figure \ref{fig:modelvar} we consider the
effect of the mass-loading factor on the abundance tracks.  Constant
values of $\eta=0$ (no outflows) and $\eta=10$ (high outflows) only
affect the behavior at high metallicity.  A time-varying model in
which $\eta(t)=10\, (t+0.1\ {\rm Gyr})^{-1}$ is also shown, and
produces a very similar track as the default model.  This strongly
varying $\eta$ is roughly equivalent to $\eta \propto M_\ast^{-0.5}$,
where $M_\ast$ is the model galaxy stellar mass in the adopted model.

The lower right panel of Figure \ref{fig:modelvar} shows additional
model variations, including a model with closed box chemical evolution
(no outflows or inflows), a shift in the onset of SNe Ia from 0.1 to
0.2 Gyr, and a model in which the gas inflow rate is an exponentially
increasing function of time ($M_g(t)=e^{t/1\, {\rm Gyr}}$).  All of
these variations produce modest shifts in the abundance tracks.  We
therefore conclude that none of them are able to mimic the
non-monotonic behavior produced by SFE variations.

Our default model is meant to be illustrative.  Unsurprisingly, it
does not reproduce other aspects of the data.  In particular, the
predicted age-metallicity relation is too shallow (see Figure
\ref{fig:model2}) and the onset of the transition to high star
formation efficiency is too late.  Furthermore, the distribution
function (MDF) becomes shallower at [Fe/H] $>-1$, in contrast to the
data, which becomes steeper.  These two shortcomings are likely
related.  For example, a time-dependent inflow rate will generally
have a small effect on the [$\alpha$/Fe]-[Fe/H] tracks \citep[see
Figure \ref{fig:modelvar}, and][]{Andrews17} but a large effect on the
age-metallicity relation and the MDF.  Simultaneously reproducing
these key constraints will provide important insights into the early
history of the Galaxy \citep[see e.g.,][for recent
examples]{Kobayashi20, Snaith22}, and will be the subject of future
work.

\subsection{Comparison to Simulations}

In this section we provide a brief comparison to simulations from the
NIHAO project \citep{Wang15}.  \citet{Buck20} showed that many
simulated NIHAO galaxies display non-monotonic abundance trends,
qualitatively similar to what we find in the H3 in-situ population.
Two striking examples are \texttt{g3.59e11} and \texttt{g7.66e11},
which have halo masses of $3.59\times 10^{11}\msun$ and
$7.66\times 10^{11}\msun$, respectively.  For these galaxies we have
computed the average stellar metallicity ([Fe/H]), oxygen-to-iron
abundance ratio ([O/Fe]), SFR averaged over 200 Myr timescales, and
the gas mass as a function of time.  These averages were computed for
the high$-\alpha$ sequences defined in \citet{Buck20}.  NIHAO does not
track Mg, so we use O as a tracer for $\alpha$ elements \citep[though
see][for an updated stellar evolution model]{Buck21}.

In Figure \ref{fig:nihao} we show the abundance patterns of these two
galaxies (left panel) and their star formation efficiencies (right
panel).  We plot these quantities as a function of the metallicity
relative to the metallicity at which the non-monotonic behavior begins
(referred to as [Fe/H]$_{\rm trans}$).  The key point is that the rise
in [O/Fe] corresponds to a sudden and significant increase in star
formation efficiency.  We take this result to corroborate our
interpretation of the data based on analytic chemical evolution
models, in which the non-monotonic behavior in [$\alpha$/Fe]
vs. [Fe/H] is due to a sudden increase in star formation efficiency.
Similar non-monotonic behavior is seen in other recent simulations
\citep[e.g.,][]{Few14, Brook20}.

The transition scale occurs at a metallicity that is somewhat too high
compared to the data ($-0.7$ and $-1.15$ compared to $-1.3$), and much
too late (6 Gyr for both simulated galaxies, compared to $\approx1$
Gyr in the data).  Moreover, the shape of the chemical evolution track
shows quantitative differences compared to the data.  Of the 35 NIHAO
galaxies presented in \citet{Buck20}, only one shows a transition at
the low metallicity seen in the data (\texttt{g5.31e11}).  This result
is in agreement with \citet{Belokurov22}, who find that the transition
between hot and cold kinematics occurs too late in the Latte
\citep{Wetzel22} and Auriga \citep{Grand18a} simulations.  For the
NIHAO simulations, mergers appear to be the main driver of the
enhanced star formation efficiency and non-monotonic abundance
behavior.  However, given that the timing is very different between
simulations and data, we reserve judgement on the physical mechanism
responsible for the transition in the data to future work.

%--------------------------------------------------------%
%--------------------------------------------------------%
%--------------------------------------------------------%

\vspace{1cm}

\section{Discussion}
\label{s:disc}

\subsection{Caveats \& Limitations}
\label{s:caveats}

Before placing these results in context, we highlight several caveats
and limitations to the present analysis.  

The H3 window function is uniform and nearly complete at
$|b|>30^\circ$ and Dec. $>-20^\circ$.  Existing H3 data does not reach
lower Galactic latitudes, and the parallax selection removes most
stars with heliocentric distances of $\lesssim 1$ kpc.  For the
high$-\alpha$ population studied here, which has a scale-height of
$0.6-1$ kpc \citep[e.g.,][]{Bovy12}, the lack of in-plane coverage of
H3 is unlikely to bias our results, but we caution that any systematic
population variation between $|Z|<1$ kpc and $|Z|>1$ kpc would not be
reflected in our results.

Our abundance determinations are based on plane-parallel 1D
atmospheres and assumed LTE.  Furthermore, our $\alpha$ abundances are
primarily derived from the \ionn{Mg}{i} triplet at 5167, 5173,
5183\AA, which is very strong and sensitive to the structure of the
upper stellar atmospheres \citep[e.g.,][]{Bergemann17a}.
\citet{Cargile20} showed that the H3 abundance scale (in both [Fe/H]
and [$\alpha$/Fe]) agrees well with literature values for globular and
open clusters spanning the metallicity range [Fe/H]$\approx-2.3$ to
$+0.0$.  Furthermore, as we show in Appendix \ref{s:comp}, the trends
we identify in abundance space are also seen in the APOGEE data.  This
comparison gives added confidence to our use of the \ionn{Mg}{i}
triplet for abundance determination.  Furthermore, the comparison
between the in-situ and accreted sequences in Figure \ref{fig:chem} --
which are on the same abundance scale -- shows that there is a feature
in the in-situ abundances at [Fe/H]$\approx-1.3$ that does not appear
in the accreted sample.  This feature therefore cannot be due to
deficiencies in the H3 abundance analysis.  Nonetheless, we caution
that the detailed quantitative behavior in abundance space is
sensitive to NLTE and 3D atmosphere effects \citep[see
e.g.,][]{Bergemann17b}, and the overall behavior of the [$\alpha$/Fe]
vs. [Fe/H] sequence at low metallicity differs across the literature
\citep[compare][]{Gratton03, Cayrel04, Ruchti11, Bensby14}.  Our
quantitative interpretation of the detailed abundance behavior at low
metallicity must therefore be treated with care.

The most important observational caveat in this work is the separation
of in-situ and accreted stars at low metallicity.  Despite decades of
effort, this task remains challenging \citep[see e.g.,][]{Buder22}.
Following the work of \citet{Hawkins15}, \citet{Belokurov22} use
[Al/Fe] to separate in-situ and accreted populations.  While this
selection provides good separation at [Fe/H] $\gtrsim-1.2$, at lower
metallicities the in-situ population overlaps in chemical space with
the accreted stars, in both the diagnostic [Al/Fe] and [Mg/Mn]
abundance ratios, as we show in Appendix \ref{s:chemaccr} \citep[and
also discussed in][]{Horta21, Belokurov22}.  This means that it will
be challenging to identify in-situ stars at low metallicity purely on
the basis of these proposed diagnostic abundance ratios.  However, our
approach in this work, which employs kinematic selection, is also
problematic.  In-situ stars on highly radial or retrograde orbits,
which exist at low metallicity \citep{Belokurov22}, will be missed by
our selection. Our results related to the kinematic behavior of the
in-situ population (Figure \ref{fig:kin}) must therefore be
interpreted with care, as we discuss in Section \ref{s:mainres}.  A
path forward may involve exploiting the predicted spatial variation in
the accreted and in-situ populations with height from the Galactic
plane in order to decompose these two populations, at least in a
statistical sense.

In order to interpret the chemical behavior of the in-situ population
we relied on an analytic chemical evolution model.  The chemical
evolution of a system is very complex and challenging to model as
there are many relevant physical processes, each with significant
uncertainties (including stellar yields, Type Ia delay time
distribution, inflow and outflow rates, mass-loading, mixing, etc.).
Within the one-zone framework considered in Section \ref{s:chemevol},
we undertook a wide exploration of the parameter space and could not
identify another physical mechanism besides a change in SFE that would
reproduce the observed non-monotonic behavior in [$\alpha$/Fe]
vs. [Fe/H].

The most obvious alternative way to produce the non-monotonic behavior
is with metallicity-dependent CC yields \cite[e.g.,][]{Heger10,
  Vincenzo16}.  However, we do not see any sign of this behavior in
our accreted star sample (which shows a smooth decline in
[$\alpha$/Fe] over the range $-2\lesssim$ [Fe/H] $\lesssim-0.5$), and
the variable SFE model reproduces the distinctive shape of the
observed trend remarkably well (Figure \ref{fig:model}).  Further
high-precision studies of large samples in this metallicity range
would be valuable for detecting or ruling out metallicity-dependent
yields, which should be governed by stellar astrophysics with little
dependence on galactic environment.  Conversely, measuring other
elements with large contributions from SN Ia or from AGB stars could
be a useful way of verifying that the early decline of [$\alpha$/Fe]
is indeed a result of a time-delayed enrichment source.

\subsection{The High$-\alpha$ Disk from Beginning to End}

We begin with a summary of the main results.  We identified a sample
of in-situ stars by selecting prograde orbits with eccentricities
$e<0.8$.  We argue that this selection results in a relatively pure
but incomplete sample of in-situ stars at low metallicity.  The
high$-\alpha$ subset of these stars forms a continuous sequence in
[$\alpha$/Fe] vs. [Fe/H].  This sequence is non-monotonic, showing
inflections at [Fe/H] $\approx-1.3$ and $\approx-0.7$.  Both analytic
chemical evolution models and realistic hydrodynamic simulations show
that this non-monotonic behavior is a reflection of a substantial
increase in star formation efficiency (SFR$/M_{\rm gas}$).  The
kinematics of the high$-\alpha$ in-situ population are hot with small
net rotation ($V/\sigma\lesssim1$) at [Fe/H] $\lesssim-1$ and become
increasingly well-ordered and cold at higher metallicities.  Main
sequence turnoff-based ages show stars at [Fe/H] $\lesssim-1.3$ are
uniformly very old ($\approx13$ Gyr), while more metal-rich stars span
a range of ages ($\approx8-12$ Gyr).

\begin{figure}[!t]
\center
\includegraphics[width=0.5\textwidth]{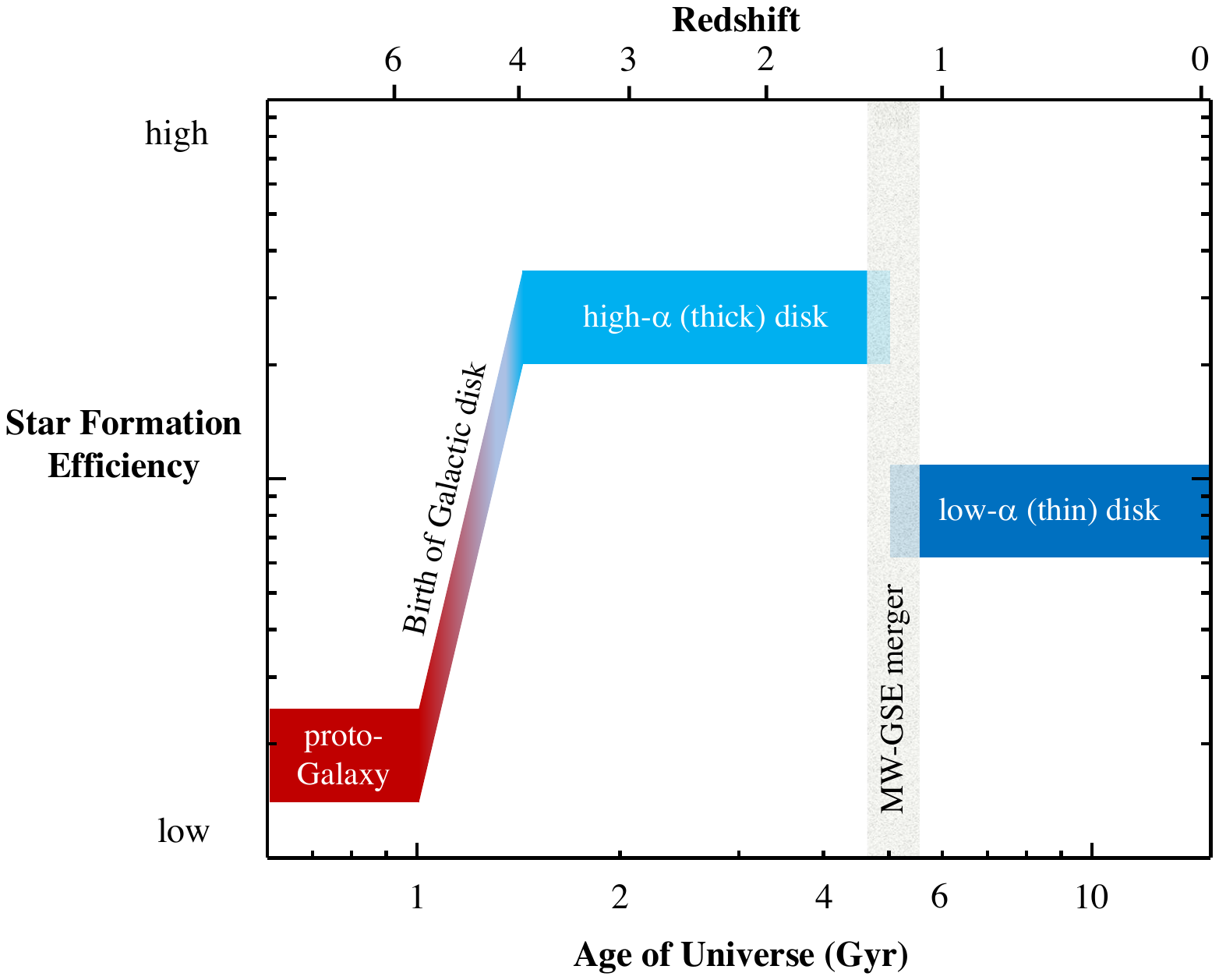}
\vspace{0.1cm}
\caption{Schematic overview of Galactic star formation efficiency
  (SFE) over cosmic time.  The colors indicate the kinematic state of
  the forming component (red $=$ hot, blue $=$ cold).  At early times
  the proto-Galaxy has low SFE and is kinematically hot.  After a
  period of $\sim1$ Gyr, the SFE rapidly increases and the kinematics
  become much colder.  This marks the beginning of the high$-\alpha$
  disk phase.  At $z\sim1$ the GSE finishes merging with the Galaxy.
  This event truncates the high$-\alpha$ disk, creates the in-situ
  halo via kicked up high$-\alpha$ disk stars, and initiates the epoch
  of low$-\alpha$ disk formation.  In this final phase, the kinematics
  are colder and the SFE lower than the high$-\alpha$ disk
  \citep{Nidever14}.  Epochs are approximate.  See text for details. }
\label{fig:cartoon}
\end{figure}

The fact that the abundance pattern forms a continuous sequence from
[Fe/H] $\approx-2.5$ to $\approx0$, combined with the observed
age-metallicity relation, supports a picture in which the
low-metallicity prograde stars are both genuinely in-situ and
represent the earliest epochs of the Galaxy.  The continuity in
abundance space argues against an accretion origin for the
low-metallicity stars \citep[cf.][]{Abadi03, Carollo19}.  This
non-monotonic abundance behavior is not unique to the Galaxy
\citep[e.g.,][]{Hendricks14, Nidever20, Hasselquist21}, and suggests
that abrupt star formation efficiency variations are common.

Observations are providing an increasingly sharp view of the
high$-\alpha$ disk from the earliest epochs to the present day.
However, the extant data do not uniquely determine the physical origin
and evolution of this Galactic component.  Here we sketch one
plausible scenario that is consistent with the existing data.  We
propose that the ancient, metal-poor in-situ stars represent the
earliest epoch of the proto-Galaxy.  During this ``simmering phase'',
the star formation efficiency was low and relatively few stars formed.
After $\approx1$ Gyr of simmering, at a critical metallicity of [Fe/H]
$\approx-1.3$, the Galaxy underwent a dramatic transition to a
``boiling phase'', in which the star formation efficiency
substantially increased, the number of stars increased, and the
kinematics became increasingly cold and disk-like \citep[][refer to
this as the ``spin-up'' phase]{Belokurov22}.  This transition marks
the birth of the Galactic disk.  The boiling phase -- the epoch of the
high$-\alpha$ disk -- lasted for $\approx3-4$ Gyr (from $z\approx4$ to
$z\approx1$), at which time a merger with GSE created the in-situ halo
from dynamically-heated disk stars, terminated the high$-\alpha$ disk
phase, and initiated growth of the low$-\alpha$ disk
\citep[e.g.,][]{Bonaca20, Belokurov20}.  The low-$\alpha$ disk is
kinematically colder and consistent with having formed at lower star
formation efficiency than the high$-\alpha$ disk
\citep[e.g.,][]{Nidever14}.  The development of a stable hot gaseous
halo may also have occurred at $z\sim1$ and regulated the delivery of
gas to the star-forming disk, enabling the formation of a
kinematically cold disk \citep[e.g.,][]{Yu21, Stern21, Gurvich22,
  Hafen22}.  Perhaps it was the merger with GSE that resulted in the
formation of the Milky Way's hot halo.  A schematic illustration of
these key phases is shown in Figure \ref{fig:cartoon}.  In a
single-zone model, resetting the low$-\alpha$ disk to low metallicity
requires dilution by a large influx of gas
\citep[e.g.,][]{Chiappini97}, but this requirement may be softened in
multi-zone models that combine a Galactic metallicity gradient with
radial migration \citep[e.g.,][]{Schonrich09, Minchev13, Loebman16,
  Sharma21, Johnson21a}.

We can place a limit on the mass of the proto-Galaxy by noticing that
stars at [Fe/H] $<-1.3$ comprise only 1\% of the high$-\alpha$ in-situ
population (Figure \ref{fig:mdf}).  Adopting a thick disk stellar mass
of $6\times10^9\,\msun$ \citep{Bland-Hawthorn16}, and equating the
thick and high-$\alpha$ disk leads us to infer a stellar mass of the
proto-Galaxy at $z\approx4$ of $\sim 6\times 10^7\,\msun$.  Such
objects should be within range of upcoming {\it James Webb Space
  Telescope (JWST)} programs \citep[e.g.,][]{Williams18}.

As noted in Section \ref{s:caveats}, it is challenging to identify
in-situ stars at low metallicity.  Our approach here relied on
kinematic separation.  \citet{Belokurov22} selected in-situ stars
based on [Al/Fe] from APOGEE.  They identified a significant change in
the kinematic behavior of in-situ stars at [Fe/H] $\approx-1$, in good
agreement with what we find here, which they interpret as the period
of disk spin-up.  Belokurov \& Kravtsov lacked empirical age
estimates, which we employ here to identify the epoch of this
transition.  Furthermore, our kinematic selection of in-situ stars
enabled identification of such stars to much lower metallicity than in
the Belokurov \& Kravtsov sample ([Fe/H] $\approx-2.5$ compared to
$-1.5$), which allowed us to identify a non-monotonic abundance
pattern and its interpretation in terms of star formation efficiency
variations.

What initiated the transition from simmering to boiling $\approx13$
Gyr ago?  Analysis of the NIHAO simulations suggest that mergers can
induce a rapid increase in star formation efficiency.  At high
redshift mergers are common \citep{Fakhouri10}, and so a merger could
have been responsible for this transition.  In this picture, the
kinematic disorder observed at [Fe/H] $\lesssim-1$ could mark the
effect of a merger (or mergers) at this epoch.  Another possibility
lies in the very rapid increase in galaxy mass and SFR at early times,
when a galaxy transitions from essentially zero SFR to a SFR of
$\sim1-10\, \msun\, {\rm yr}^{-1}$ within $\sim1$ Gyr
\citep[e.g.,][]{Renaud21, Belokurov22}.  Perhaps in this early, rapid
growth phase, the ratio of SFR and gas mass accretion rates are out of
sync so that the star formation efficiency experiences a rapid
increase.  An alternative idea advanced by \cite{Mason24}, motivated
by the EAGLE simulations, is that the effective star formation
efficiency is low at early times because metals from supernovae are
vented and mixed into the gaseous halo, which is not forming stars,
and the efficiency grows when metals are trapped in the star-forming
ISM.  \cite{Chen24} show that non-monotonic behavior of [$\alpha$/Fe] can
be triggered by rapid changes in gas infall rate rather than
efficiency.  We encourage further exploration of simulations at these
early epochs to understand this important phase in the evolution of
our Galaxy.

The results presented in this work, in combination with recent results
from \citet{Belokurov22}, mark a new era in our ability to observe in
the archeological record the very earliest phases of the Galaxy.
These and future observations within the Galaxy should provide
powerful constraints on models of the high-redshift universe that will
be complementary to forthcoming observations from {\it JWST}.

%--------------------------------------------------------%

\acknowledgments 

CC thanks Vasily Belokurov and Andrey Kravtsov for very helpful
conversations and feedback on an earlier draft.  CC and PC acknowledge
support from NSF grant NSF AST-2107253.  DHW and JWJ acknowledge
support from NSF grant NSF AST-1909841.  RPN acknowledges an Ashford
Fellowship granted by Harvard University. TB acknowledges support by
the European Research Council under ERC-CoG grant CRAGSMAN-646955.
YST acknowledges financial support from the Australian Research
Council through DECRA Fellowship DE220101520.  We thank the
Hectochelle operators and the CfA and U. Arizona TACs for their
continued support of the H3 Survey.

Observations reported here were obtained at the MMT Observatory, a
joint facility of the Smithsonian Institution and the University of
Arizona.  This paper uses data products produced by the OIR Telescope
Data Center, supported by the Smithsonian Astrophysical
Observatory. The computations in this paper were run on the FASRC
Cannon cluster supported by the FAS Division of Science Research
Computing Group at Harvard University.  This research made use of the
NumPy \citep{Harris20}, pynbody \citep{pynbody} and tangos
\citep{tangos} package to analyze the simulations. TB gratefully
acknowledges the Gauss Centre for Supercomputing
e.V. (www.gauss-centre.eu) for funding this project by providing
computing time on the GCS Supercomputer SuperMUC at Leibniz
Supercomputing Centre (www.lrz.de) under the project number pn29mo.

This work has made use of data from the European Space Agency (ESA)
mission {\it Gaia} (\url{https://www.cosmos.esa.int/gaia}), processed
by the {\it Gaia} Data Processing and Analysis Consortium (DPAC,
\url{https://www.cosmos.esa.int/web/gaia/dpac/consortium})
\citep{GaiaEDR3}. Funding for the DPAC has been provided by national
institutions, in particular the institutions participating in the {\it
  Gaia} Multilateral Agreement.  Funding for the Sloan Digital Sky
Survey IV has been provided by the Alfred P. Sloan Foundation, the
U.S.  Department of Energy Office of Science, and the Participating
Institutions. SDSS-IV acknowledges support and resources from the
Center for High Performance Computing at the University of Utah. The
SDSS website is \url{www.sdss.org}.

%--------------------------------------------------------%

%\bibliography{../master_refs}

\begin{appendix}

\section{Comparison to APOGEE}
\label{s:comp}

\begin{figure*}[!t]
\center
\includegraphics[width=0.95\textwidth]{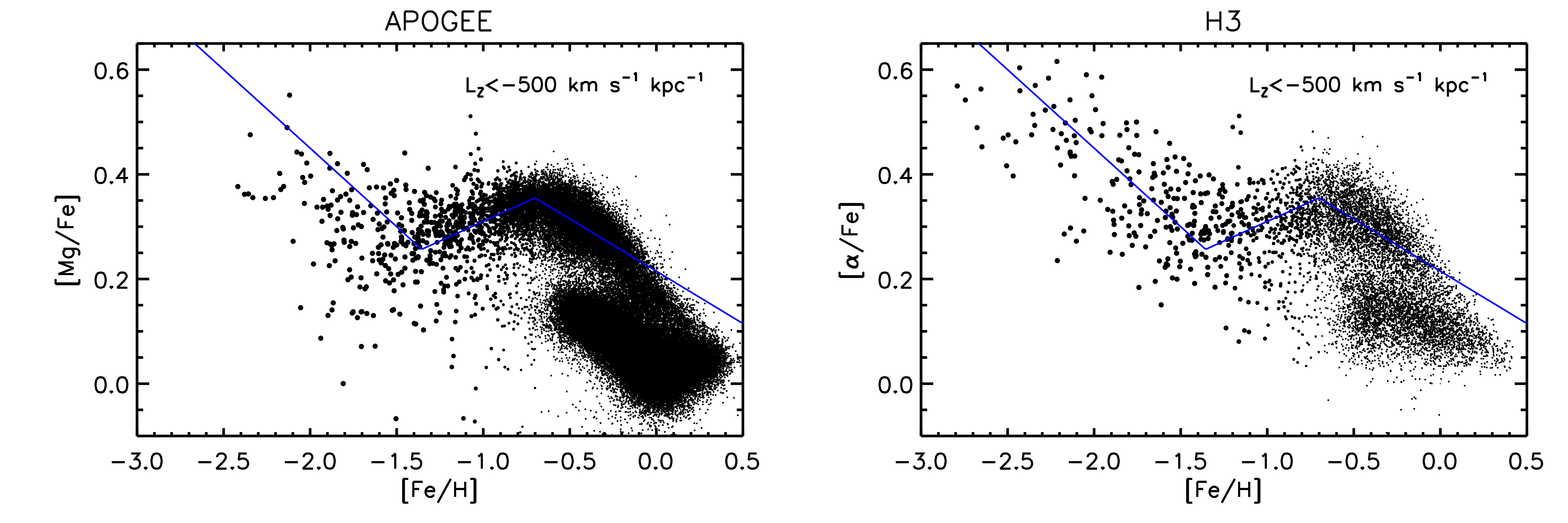}
\vspace{0.1cm}
\caption{Comparison between APOGEE (left panel) and H3 (right panel)
  data for prograde stars ($L_Z<-500$ km s$^{-1}$ kpc$^{-1}$).  The
  blue line is intended to guide the eye.  The overall trend of the
  high$-\alpha$ sequence is the same in the two surveys, although the
  APOGEE data are offset to lower [Mg/Fe] at [Fe/H] $\lesssim-1.5$.
  In H3 only the global [$\alpha$/Fe] is measured, although it is
  primarily sensitive to Mg.  Therefore, in the comparison to APOGEE
  we plot [Mg/Fe] for that survey.  In both panels the symbol size is
  inversely proportional to metallicity in order to draw attention to
  the low-metallicity sequence.}
\label{fig:apo}
\end{figure*}

In this Appendix we compare the H3 chemistry of in-situ (prograde)
stars to stars in APOGEE DR17 \citep{APODR17}.  For distances and
angular momenta we use the AstroNN value-added catalog from
\citet{Leung19}.  Following \citet{Belokurov22}, we restrict the
APOGEE catalog to log $g<3.0$ and remove stars associated with the
Magellanic Cloud program and any cluster-related program.  We also
remove stars with potentially problematic measurements via the flags
\texttt{STAR\_BAD}, \texttt{TEFF\_BAD}, \texttt{LOGG\_BAD},
\texttt{PERSIST\_HIGH}, \texttt{VERY\_BRIGHT\_NEIGHBOR},
\texttt{PERSIST\_JUMP\_POS}, \texttt{PERSIST\_JUMP\_NEG}, or
\texttt{SUSPECT\_RV\_COMBINATION}.  We also require SNR$>200$, {\it
  Gaia} EDR3 parallax SNR$>10$, and agreement between the {\it Gaia}
parallax and the AstroNN distances to 50\%.  To enable a fair
comparison to the H3 sample we restrict the APOGEE sample to
$6<R_{\rm gal}<10$ kpc.  Finally, to focus on the in-situ (prograde)
population we select stars with $L_Z<-500$ km s$^{-1}$ kpc$^{-1}$ (the
AstroNN coordinate system has a different sign than ours; we have
converted their system to ours for consistency).  With these cuts the
final sample contains 74,806 stars.  For this sample the median formal
uncertainty on [Mg/Fe] and [Fe/H] are $\approx0.01$.

Figure \ref{fig:apo} compares the chemistry of prograde APOGEE and H3
stars.  For the former we use [Mg/Fe] since the H3 spectral window is
mostly sensitive to Mg among the $\alpha$ elements.  APOGEE does not
take atomic diffusion into account, and so in this figure we use the
surface abundances reported for H3 stars, as opposed to the initial
composition.  The behavior of the high$-\alpha$ sequences are nearly
identical for [Fe/H] $\gtrsim-1.3$.  This agreement provides
confirmation that both abundance scales are reliable, as the analysis
pipelines of H3 and APOGEE are quite different (although with common
elements such as the assumption of LTE).  At lower metallicities the
APOGEE data shows an increase in [Mg/Fe] as also seen in the H3 data.
However, the normalization is lower by $\approx0.1$, or,
alternatively, the metallicity scale of APOGEE and H3 are offset by
$\approx0.1$ at [Fe/H] $\lesssim= -1.5$ \citep[see][]{Cargile20}.
Overall, this comparison provides independent confirmation of the
abundance trends reported in this paper.

\section{On the Use of Chemistry to Identify Accreted Stars}
\label{s:chemaccr}

\begin{figure*}[!t]
\center
\includegraphics[width=0.95\textwidth]{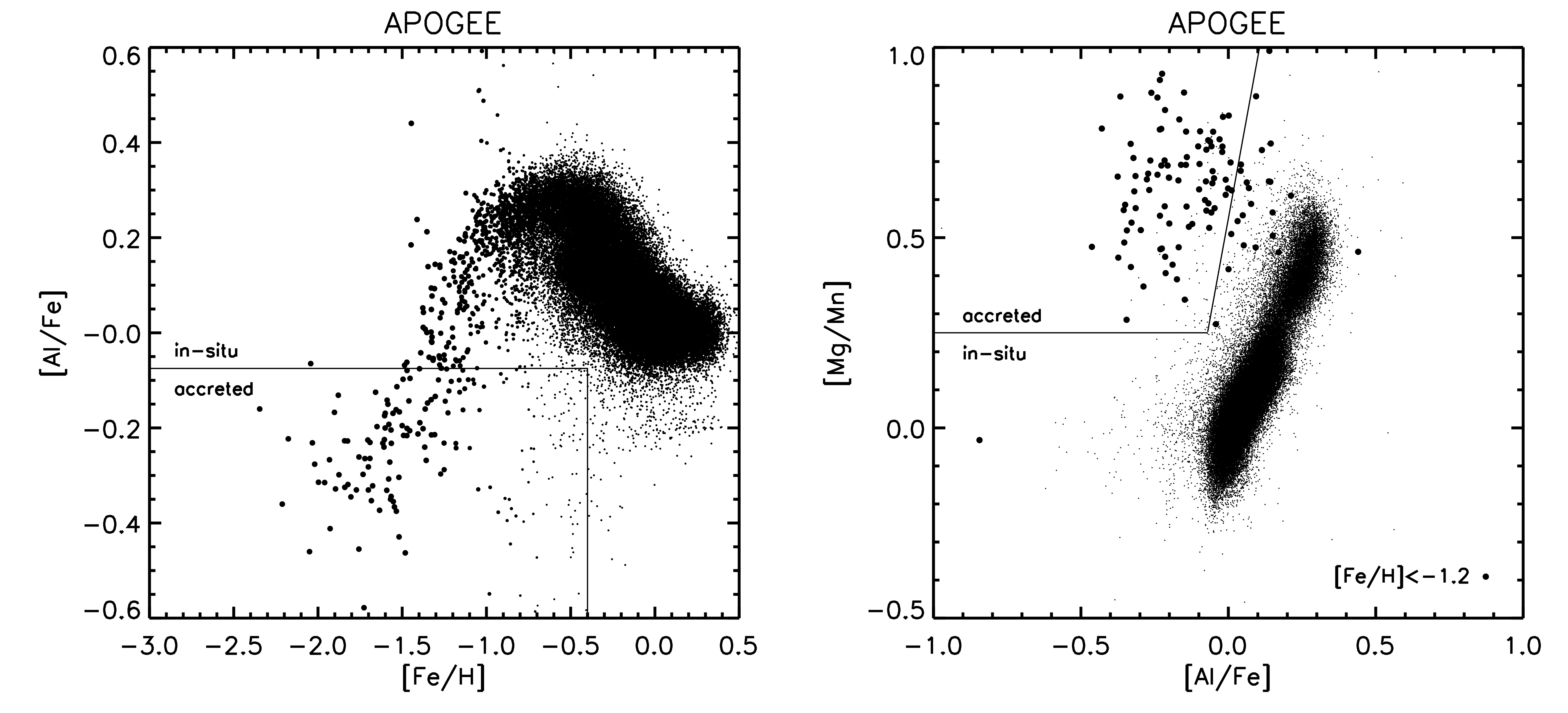}
\vspace{0.1cm}
\caption{Chemistry of stars in APOGEE with $L_Z<-10^3$ km s$^{-1}$
  kpc$^{-1}$.  Such stars have a median eccentricity of $e\approx0.2$,
  i.e., they are on nearly circular prograde orbits and so are very
  unlikely to be accreted.  Left panel: [Al/Fe] vs. [Fe/H].  This
  diagram was recently used by \citet{Belokurov22} to separate
  accreted and in-situ stars, with the former presumed to occupy the
  lower left region indicated by the box.  In this same region we also
  clearly see metal-poor in-situ stars.  Symbol size is inversely
  proportional to metallicity in order to draw attention to the
  low-metallicity sequence.  Right panel: [Mg/Mn] vs. [Al/Fe].  This
  diagram has been proposed as an efficient way to separated accreted
  and in-situ stars, with the former occupying the upper left region
  indicated by the lines \citep[see e.g.,][]{Hawkins15, Horta21}.  In
  this panel stars with [Fe/H] $<-1.2$ are indicated with large
  symbols.  It is clear that the in-situ, metal-poor stars have
  similar chemical make-up as accreted stars, at least with regards to
  [Al/Fe] and [Mg/Mn].  This method of selecting accreted stars should
  therefore be treated with caution at low metallicity.}
\label{fig:alfe}
\end{figure*}

\citet{Hawkins15} proposed the use of the abundance ratios [Al/Fe] and
[Mg/Mn] to separate accreted and in-situ stars.  A number of authors
have since used some combination of these abundance ratios (and
[Fe/H]) to separate accreted from in-situ stars \citep[e.g.,][]{Das20,
  Horta21, Belokurov22}.  In this section we briefly revisit these
selections in the context of metal-poor in-situ stars.

Figure \ref{fig:alfe} shows two projections of abundance space:
[Al/Fe] vs. [Fe/H] (left panel) and [Mg/Mn] vs. [Al/Fe] (right panel).
The boxes mark regions where accreted stars are supposed to reside
\citep[cf.][]{Das20, Horta21, Belokurov22}.  However, in this figure
we plot only APOGEE stars with significant prograde motion,
specifically $L_Z<-10^3$ km s$^{-1}$ kpc$^{-1}$.  Such stars have an
average eccentricity of 0.15 -- i.e., they are on nearly circular
orbits and so are very unlikely to be accreted stars.  In the left
panel one sees that nearly every prograde star at [Fe/H] $<-1.5$
resides in the accretion selection box.  In the right panel stars with
[Fe/H] $<-1.2$ are displayed as large symbols; one clearly sees that
the overwhelming majority of such stars reside in the accretion
selection box.  The contamination of in-situ stars in the accretion
selection box at low metallicity was also pointed out by
\citet{Horta21} and \citet{Belokurov22}.

Figure \ref{fig:alfe} shows that the use of these chemistry-based
selections will incorrectly assign genuine low-metallicity, in-situ
stars to the accreted population.  The contamination of the resulting
accretion-selected stars is likely to be low, as low-metallicity
in-situ stars are rare.  However, the selection will bias (truncate)
the low-metallicity in-situ population.  We therefore urge caution
when using a chemistry-based separation between in-situ and accreted
stars if one is interested in low-metallicity in-situ stars.

\end{appendix}

\end{CJK*}

\end{document}